\documentclass[aps,
prx,
superscriptaddress,
reprint,
longbibliography,
floatfix]{revtex4-2}

\usepackage{amsmath,amssymb,amsfonts}
\usepackage{graphicx}
\usepackage{color}
\usepackage{xcolor}
\usepackage{physics}
\usepackage{bm}
\usepackage{enumitem}

\usepackage[colorlinks=true,unicode]{hyperref}

\usepackage{mathtools}
\usepackage{booktabs}
\usepackage{needspace}
\usepackage[T1]{fontenc}

\newcommand{\Pacc}{P_{\mathrm{acc}}}
\newcommand{\Plog}{P_{\mathrm{L}}}

\newcommand{\Cfour}{\mathrm{C4}}
\newcommand{\GKP}{\mathrm{GKP}}

\newcommand{\MN}{\mathrm{MN}}

\newcommand{\act}{a}

\usepackage{tikz}
\usetikzlibrary{arrows.meta}
\newcommand{\red}{\color{red!70!black}}
\definecolor{darkrailred}{RGB}{139,0,0}
\definecolor{darkrailblue}{RGB}{0,0,139}
\definecolor{paleyellow}{RGB}{255,245,170}
\definecolor{palegreen}{RGB}{185,230,180}
\definecolor{pinkhighlight}{RGB}{255,190,190}
\tikzset{
  horizontal rail/.style={draw=darkrailred,line width=1.1pt,line cap=round},
  vertical rail/.style={draw=darkrailblue,line width=1.1pt,line cap=round},
  cell arrow 1/.style={draw=black,-{Stealth[length=3pt,width=2.4pt]},shorten >=2.2pt,line width=.65pt},
  cell arrow 2/.style={draw=gray,-{Stealth[length=3pt,width=2.4pt]},shorten >=2.2pt,line width=.65pt},
  mode node/.style={circle,draw=black,line width=.65pt,inner sep=0pt,minimum size=5.5pt},
  connected highlight/.style={draw=none,fill=pinkhighlight,rounded corners=7pt},
  zigzag region/.style={draw=pinkhighlight,line width=14pt,line cap=round,line join=round},
  node shade/.style={draw=none,fill=pinkhighlight},
  pale yellow node/.style={mode node,fill=paleyellow},
  pale green node/.style={mode node,fill=palegreen},
  selected label/.style={font=\sffamily\small,text=black},
  time label/.style={font=\sffamily\large,text=black!55},
  legend text/.style={font=\small,text=black,anchor=west,inner sep=0pt}
}

\newcounter{criterion}[section]
\renewcommand{\thecriterion}{\thesection.\arabic{criterion}}
\newenvironment{criterionbox}[1][]{%
 \par\Needspace{6\baselineskip}%
 \refstepcounter{criterion}%
 \par\medskip\noindent\textbf{Operational criterion \thecriterion. #1}\quad
}{\par\medskip}

\newcommand{\FinalRecursiveSchedule}{(q_2,q_3,q_4,q_5)=(1,0.70,0.960,1.0)}

\hypersetup{linkcolor=black,citecolor=black,urlcolor=black,
  pdftitle={Modular optical fault-tolerant quantum computation via an interface between verified resources and online logical operations},
  pdfauthor={Kosuke Fukui and coauthors},pdfsubject={Modular optical fault-tolerant quantum computation}}

\begin{document}
\raggedbottom

\title{Modular optical fault-tolerant quantum computation\\via an interface between verified resources and online logical operations}

\author{Kosuke Fukui}
\author{Takaya Matsuura}
\author{Yohji Chin}
\author{Shohei Kiryu}
\author{Takaya Hoshi}
\author{Kan Takase}
\author{Warit Asavanant}
\affiliation{OptQC Corp., 3-28-13 Nishi-Ikebukuro, Toshima-ku, Tokyo 171-0021, Japan}

\begin{abstract}
We introduce a factory--processor interface architecture for modular optical fault-tolerant quantum computation that separates code-dependent encoded-resource preparation from reusable online logical processing on a common Gaussian optical processor. Beyond encoded resources, the interface specifies logical outputs, code-to-mode mapping, measurement roles, frame updates, verification records, and resource-injection boundaries. As a concrete realization, we combine an offline verified Bell-resource factory based on a recursively concatenated $C_4$ code with physical Gottesman--Kitaev--Preskill (GKP) qubits with an online time-domain measurement-based quantum computation (MBQC) processor. Under the specified additive-displacement-noise model and fixed postselection policy, simulations support the finite-level logical Bell-error ordering $P_5<P_4<P_3<P_2$ at 12.25 dB of effective GKP displacement-noise quality, both with and without the prescribed additional post-injection displacement noise. At 15.6 dB and $L=2$, a separate fixed-sample evaluation gives one-sided 95$\%$ Clopper--Pearson upper bounds below $10^{-5}$ for both conditions under the binomial model. The interface supports address-tracked injection, transversal CNOT and logical Bell-measurement constructions, and logical SWAP and Hadamard constructions with tracked code permutations. Given advance resource preparation and sufficient routing parallelism, increasing the concatenation level expands parallel encoded-resource width without increasing the protected online sequence length of these constructions. These results identify a modular route to lower logical error rates while retaining a common Gaussian MBQC operation set with shallow online processing.
\end{abstract}

\maketitle

\section{Introduction}

Quantum algorithms motivate computing systems that can sustain long quantum computations despite imperfect physical operations~\cite{Shor1994Algorithms,GidneyEkera2021RSA}. Fault-tolerant quantum computation (FTQC) addresses this requirement through encoding, quantum error correction (QEC), and protected logical operations, with resource overhead depending on both the code and its implementation~\cite{Fowler2012SurfaceCode}.
Hardware-aware FTQC architectures couple noise, resource supply, connectivity, and logical operations across superconducting and bosonic~\cite{Guillaud2019RepetitionCat,Kubica2023Erasure}, trapped-ion and neutral-atom~\cite{Nickerson2014Cells,Ramette2022Cavity,Cong2022Rydberg,Cai2023Looped}, and semiconductor platforms~\cite{Jones2012Layered}. They motivate interfaces that specify more than code performance alone.

Optical FTQC poses a related question in measurement-based quantum computation (MBQC), where entangled resources, measurements, and feedforward realize logical operations~\cite{Raussendorf2001OneWay,raussendorf2003measurement,Raussendorf2007Topological}. Continuous-variable (CV) clusters generated from squeezed light and linear optics~\cite{Weedbrook2012CVReview,Takeda2019PhotonicReview,fukui2022building,Menicucci2006CVCluster} can be multiplexed in time; large cluster-state experiments and programmable Gaussian processing demonstrate this platform~\cite{Yokoyama2013UltraLargeCluster,Yoshikawa2016OneMillionModes,Asavanant2019TDM2D,Larsen2019Deterministic2D,Yokoyama2025FullStackQRL}. The quad-rail lattice (QRL) arranges time-domain-multiplexed (TDM) modes into macronodes whose measurement bases implement Gaussian processing and routing~\cite{alexander2016flexible,MenicucciQRL2020,Walshe2021Macronode}.
On this Gaussian platform, non-Gaussian GKP resources provide qubit encoding and analog displacement-syndrome information for higher-level decoding~\cite{GKP2001,Menicucci2014FTCVCluster,walshe2020continuous}. The input values quoted in decibels, including $12.25$ and $15.6$ dB, denote an effective GKP displacement-noise parameter rather than measured optical-source or cluster squeezing (Sec.~\ref{sec:gkp_interface}). Recent propagating-state and integrated-source demonstrations advance optical GKP supply~\cite{Konno2024Science,Larsen2025Nature}. Universal FTQC also requires non-Clifford resources; measurement-induced cubic operations exemplify related optical techniques~\cite{Yukawa2013CubicNonlinearity,Miyata2016AdaptiveCubicGate,Marek2018NonlinearQuadratureGates,Konno2021NonlinearSqueezing,Sakaguchi2023NonlinearFeedforward}.

Optical FTQC proposals span GKP topological and concatenated coding and more general CV noise~\cite{Fukui2018Analog,Fukui2020Temporal3D,Yamasaki2020Polylog,Larsen2021FTCVMBQC,Bourassa2021Blueprint,Tzitrin2021Static,Noh2022LowOverhead,Matsuura2026GeneralNoise}, as well as interleaving, switching-free organization, and all-optical feedforward~\cite{Bombin2021Interleaving,Asavanant2023SwitchingFree,Renault2025Switchless,Hoshi2026AllOpticalTeleportation}. Complementary approaches address linear-optical and fusion-based computation~\cite{Kok2007LOQCReview,Rudolph2017WhyPhotonic,Ewert2014BoostedFusion,Bartolucci2023Fusion}, modular networking and manufacturable hardware~\cite{AghaeeRad2025ModularPhotonic,PsiQuantum2025Manufacturable}, nonlinear photonics~\cite{Yu2026PhotonicNonlinearity}, and code-agnostic linear optics~\cite{Walshe2025ArbitraryCodes}. Verified teleportation ancillas and prepared-ancilla constant-depth Clifford circuits provide earlier foundations~\cite{Knill2005,AGP2008Postselection,Zheng2020ConstantDepth}. Our focus is the operation-aware factory--processor boundary that connects such resources to their online logical use.

A central architectural challenge in optical FTQC is to specify how verified encoded resources and their associated classical records are transferred from offline preparation to online logical operations on a Gaussian processor.
A resource state and its estimated error rate alone do not fix this interface. The processor must identify the specified logical subsystem, pair physical code indices, assign measurement roles, and interpret byproducts during MBQC. 
In TDM, modes on the same rail can occupy different time bins, while a known permutation can change the output port carrying the intended state. Address and frame records must follow the quantum information, and verification outcomes and analog likelihoods must retain their meaning for the receiving decoder. When encoding, verification, or decoding changes, this boundary must continue to distinguish resource quality, operational compatibility, and execution cost. 
This factory--processor boundary differs from noisy links between separate quantum processing units investigated in Refs.~\cite{Marqversen2025FTInterfaces,Chelluri2026NoisyInterfaces}.

\begin{figure*}[t]
\centering
\includegraphics[width=\textwidth]{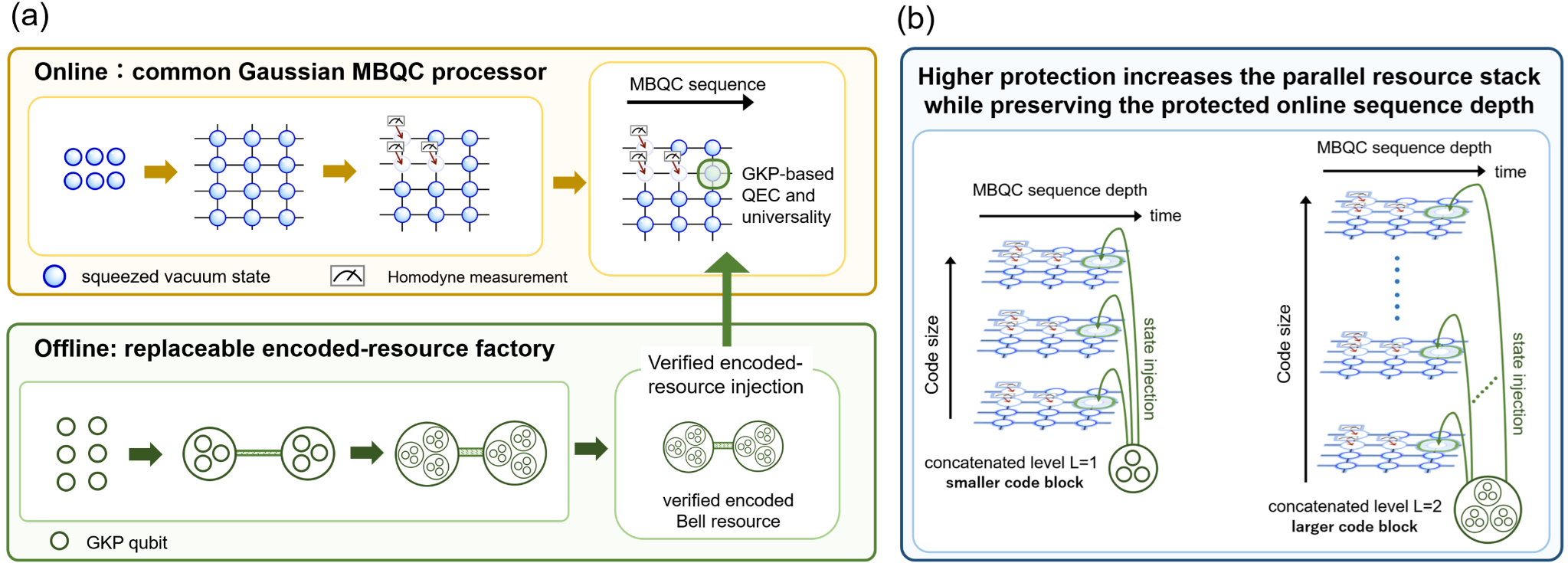}
\caption{Modular optical FTQC architecture and the verified-resource interface. (a) Encoded resources are prepared and verified in an offline factory and supplied to a common Gaussian MBQC processor. The interface associates the designated quantum resource with its mode addresses, measurement roles, frame information, and verification/reliability records; these classical records are specified in Sec.~\ref{sec:interface_contract} and are not drawn as separate wires. (b) The two illustrated concatenation levels use the same interface schema with level-dependent block sizes, addresses, and decoding. The stack represents parallel encoded resources, not additional sequential measurement rounds. An unchanged protected online sequence is conditional on advance preparation, simultaneous execution, and address-preserving routing, as discussed in Sec.~\ref{subsec:interface_execution}; it does not imply unchanged preparation time or elapsed runtime.}
\label{concept0}
\end{figure*}

We formulate a \emph{factory--processor interface architecture} comprising the designated encoded output, code-index-to-space--time addresses, measurement and frame rules, reliability records, and injection boundary. Compatibility requires matched quantum operations and classical updates for the intended logical action (Fig.~\ref{concept0}). Factories may differ in preparation, verification, decoding, logical error, acceptance, and cost while sharing the QRL Gaussian resource-state family and operation set; their address and control rules remain code dependent.
The interface is particularly relevant to \emph{transversal logical operations}: physical Gaussian measurement patterns are paired across corresponding code indices, with required code permutations and Pauli/Clifford byproducts explicitly tracked in the address and correction records. For the supported ${\rm C_4}$ Clifford constructions, this enables level-independent protected online measurement depth under advance resource preparation and sufficient address-preserving parallel capacity; transversality alone does not guarantee level-independent elapsed runtime.

For a concrete realization, we combine GKP qubits with concatenated ${\rm C_4}$ blocks used as $[[4,1,1,2]]$ data--gauge subsystem codes in a Knill-type verified Bell-resource factory. 
Both the data and gauge degrees of freedom of the ${\rm C_4}$ stabilizer block are tracked, but only the designated $D$-Bell pair is counted as the delivered output. A recursive decoder combines observable analog likelihoods, syndromes, verification outcomes, and tracked frames, building on soft-decision and concatenated-code decoding~\cite{GotoUchikawa2013Soft,Poulin2006Concatenated,Fukui2018Analog,walshe2020continuous}. Its reliability score determines binary acceptance under a schedule fixed before independent evaluation. Related resource-postselection approaches also use decoder information~\cite{Staples2026ScalablePostselection,Birchall2026Macromux}. 
The accepted resource and its associated records supply the inputs to the QRL injection and logical-operation constructions.

Two numerical results characterize the supplied resource under the specified additive-displacement-noise model. At $12.25\,\mathrm{dB}$, the fixed-schedule evaluation shows logical-error reduction from $L=2$ through $L=5$ for both specified post-injection conditions; the claim is restricted to the evaluated finite levels and grid. At $15.6\,\mathrm{dB}$ and $L=2$, an independent fixed-sample confirmation gives one-sided $95\%$ Clopper--Pearson upper bounds below $10^{-5}$ for the designated Bell-resource error, separately without and with the additional macronode-noise step. Throughout this comparison, decibels parameterize effective GKP displacement noise rather than directly measured source squeezing. The evaluated $14$-dB level-3 and $15.6$-dB level-2 points use nominal supplied-GKP counts of $3637.9$ and $288$, respectively. The within-factory comparison isolates the dependence of selected-resource cost on input quality; total-hardware and equal-input-quality benchmarking require separate accounting.

Sections~\ref{sec:model} and~\ref{sec:interface_contract} describe the architecture and interface; Sec.~\ref{sec:details} describes the factory and its decoder-based selection procedure. Section~\ref{sec:threshold} presents the principal numerical results, with statistical procedures, accounting, and auxiliary analyses in the Appendices. Section~\ref{sec:disc} concludes.

\section{Modular architecture for optical FTQC}\label{sec:model}

\subsection{Separation of the online Gaussian processor and offline encoded-resource factory}

Figure~\ref{concept0} shows an offline encoded-resource factory alongside a continuously generated Gaussian QRL processor. The factory supplies verified resources and synchronized classical records; the processor consumes them through the interface introduced in Sec.~\ref{sec:interface_contract}. As successive cluster modes are measured, quantum information propagates to subsequent time-bin modes through measurement-based teleportation. The common element is therefore the QRL resource-state family and Gaussian operation set, not fixed optical modes or a code-independent address allocation.

\subsection{MBQC platform: 2D CV cluster states}\label{sec:cv_cluster_macronode}

\begin{figure*}[t]
\centering \includegraphics[angle=0, width=1.0\textwidth]{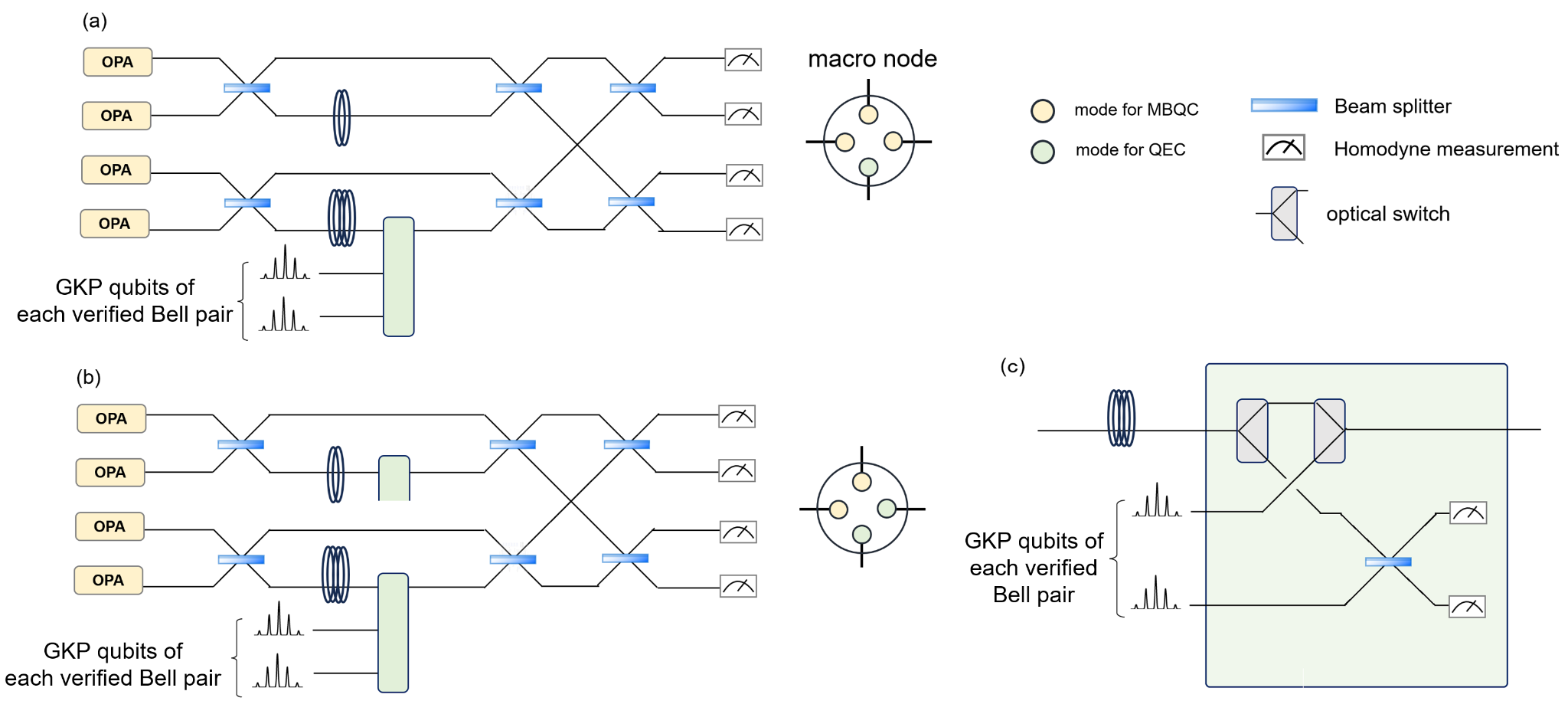} 
\caption{Schematic of the large-scale Gaussian MBQC platform based on a squeezed-vacuum 2D CV cluster. The proposed architecture employs the quad-rail lattice (QRL) with time-domain multiplexing, and Gaussian MBQC operations are implemented by homodyne measurements and classical feedforward. Fault-tolerant logical operations are supplied by verified GKP resources with higher-level qubit encoding injected into selected macronodes. Panels (a) and (b) show configurations with one and two logical-QEC/refresh ports per macronode, respectively; in this paper we consider the configuration in (a). The GKP waveforms in (a) and (b) show a matched pair of physical modes from the two encoded blocks of a verified ${\rm C_4}$/GKP Bell resource, rather than the entire encoded Bell pair. Panel (c) sketches the teleportation-based injection circuit for one such matched code index; the full encoded operation uses the corresponding physical-mode indices together with tracked classical frames and logical decoding.}
\label{MBQClayer}
\end{figure*}

We assume a large-scale two-dimensional cluster state generated from squeezed vacuum states using linear optics and time-domain multiplexing, as shown in Fig.~\ref{MBQClayer}. Specifically, we adopt the macronode protocol based on the quad-rail lattice (QRL). Nearest-neighbor entanglement between macronodes is generated by Gaussian linear-optical networks with time delays, and homodyne measurement bases determine single-mode Gaussian gates, two-mode entangling operations, and routing operations. In this paper we use the configuration in Fig.~\ref{MBQClayer}(a), with one logical-QEC/refresh port per macronode.

The two-mode QRL gadget implements direct CV SWAP through four homodyne angles~\cite{Walshe2021Macronode}. It also acts on GKP-encoded inputs and supplies the address-tracked SWAP construction of Sec.~\ref{subsec:interface_execution}.
Universal QC additionally requires appropriate non-Gaussian encoded resources (Sec.~\ref{subsec:universality}). 
The QRL implements programmable but noisy Gaussian transformations. Teleportation-based injection localizes verification and repreparation to the offline factory without restarting the Gaussian-cluster stream. 
CV ZX calculus offers a complementary process-description language~\cite{Nagayoshi2025CVZX}; here the explicit QRL model keeps physical routing and macronode layers visible.

\subsection{QEC with GKP qubits}\label{sec:gkp_interface}

Gottesman--Kitaev--Preskill (GKP) qubits provide a natural interface between CV optics and qubit-based fault tolerance. Small phase-space shifts are converted by GKP digitization into correctable residual shifts and discrete Pauli-frame errors, while Gaussian operations and homodyne measurements provide the analog information used by the higher-level decoder.

For a square-lattice GKP code with spacing $\sqrt{\pi}$, the probability that a Gaussian shift is digitized into the incorrect qubit bin is
\begin{equation}
\begin{aligned}
P_{\rm err}(\sigma)
&=
\sum_{k\in\mathbb Z}
\int_{(2k+1/2)\sqrt{\pi}}^{(2k+3/2)\sqrt{\pi}}
 d\xi\,
\frac{1}{\sqrt{2\pi\sigma^2}}
\exp\!\left[-\frac{\xi^2}{2\sigma^2}\right].
\end{aligned}
\label{eq:gkp_digitized_error_main}
\end{equation}
We use $[\hat q,\hat p]=i$ and vacuum variance $1/2$. A Gaussian displacement with variance $\sigma^2$ in either quadrature is assigned the numerical GKP displacement-noise level $s_{\rm GKP}\equiv10\log_{10}[(1/2)/\sigma^2]=-10\log_{10}(2\sigma^2)$ in decibels; for example, $s_{\rm GKP}=12$ means $12\,\mathrm{dB}$. For a numerical dB argument $s$, define $\sigma_{\rm GKP}^2(s)\equiv\tfrac12 10^{-s/10}$ and use $\sigma_{\rm GKP}^2\equiv\sigma_{\rm GKP}^2(s_{\rm GKP})$ when the operating point is understood. We use this quantity only as the equivalent noise parameter of the additive-displacement model; it neither represents directly measured source squeezing nor specifies a state-generation protocol. Unless explicitly identified as optical-source or cluster squeezing, every reported GKP displacement-noise level in decibels refers to $s_{\rm GKP}$. The homodyne residual is retained as a log-likelihood ratio (LLR) and passed to the ${\rm C_4}$ decoder~\cite{GKP2001,glancy2006error,fukui2017analog}. We denote one measured modular GKP residual by $z$; the later symbol $y_{\rm prep}$ denotes the complete multi-outcome factory record and must not be confused with this scalar residual. For an accessible residual $z$ and the alternative binary error hypotheses $b=0,1$, we fix the likelihood-ratio convention
\begin{equation}
\ell(z)=\log\frac{p(z\mid b=0)}{p(z\mid b=1)}.
\label{eq:gkp_llr_main}
\end{equation}
The two likelihoods are obtained by summing the Gaussian-displacement density over the appropriate even and odd lattice cosets. For a chosen quadrature, $b=0$ and $b=1$ label the two GKP digitization-parity hypotheses (even and odd wrapped-lattice bins), not two different physical resource types. If the class priors are unequal, the log posterior odds additionally contain $\log[P(b=0)/P(b=1)]$; a likelihood ratio must not be confused with posterior odds.

\subsection{QEC with higher-level encodings}

GKP digitization alone does not ensure a sufficiently small logical failure probability for long computations at realistic finite noise. A higher-level qubit code supplies parity constraints, repeated error detection and/or correction, and a route to stronger suppression when combined with an appropriate fault-tolerant preparation schedule. The factory--processor interface is not restricted to the particular higher-level code used below.

We use the four-qubit ${\rm C_4}$ stabilizer block $[[4,2,2]]$ in its $[[4,1,1,2]]$ subsystem interpretation: one protected information qubit $D$ is the designated data output, while the other encoded degree of freedom $G$ is treated as a tracked gauge qubit. The notation $[[n,k,r,d]]$ specifies $n$ physical qubits, $k$ protected data qubits, $r$ gauge qubits, and dressed distance $d$. After this distinction, we refer to the construction simply as the ${\rm C_4}$ code. The underlying two-qubit stabilizer space is unchanged; the chosen output/gauge roles are part of the interface contract.

A Pauli representative convention is
\begin{equation}
\begin{gathered}
\bar X_D=X_1X_2,\quad\bar Z_D=Z_1Z_3,\\
\bar X_G=X_1X_3,\quad\bar Z_G=Z_1Z_2,\\
S_X=X_1X_2X_3X_4,\quad S_Z=Z_1Z_2Z_3Z_4.
\end{gathered}
\label{eq:c4_data_gauge_main}
\end{equation}
These representatives commute with the stabilizers; the data and gauge algebras commute with each other. The accepted logical Bell pair on $D$ is the only certified output. Gauge information is propagated for parity, frame, and role tracking, not independently credited as a second output. A physical permutation that exchanges $D$ and $G$ requires an explicit output-role reassignment and does not, by itself, guarantee protection of arbitrary information initially residing in the gauge degree of freedom. Appendix~\ref{sec:concat_c4_stabilizers} supplies the full code construction.
A \emph{logical Pauli-error class} on the designated data qubit is the action of $\bar X_D^{x}\bar Z_D^{z}$, with $(x,z)\in\{0,1\}^2$, modulo irrelevant global phases, stabilizers, and gauge-equivalent representatives. We denote its four possible labels by $\mathcal A_D=\{I_D,X_D,Y_D,Z_D\}$, and the analogous gauge alphabet by $\mathcal A_G=\{I_G,X_G,Y_G,Z_G\}$; the identity class is $0$, and $\oplus$ denotes bitwise addition of their $(x,z)$ labels modulo two. A class label is an error hypothesis, not a child-resource type or physical slot. For joint data--gauge diagnostics the label space is $\mathcal A_D\times\mathcal A_G$ (16 possible pairs), without implying statistical independence between the two degrees of freedom.

\subsection{GKP analog information and soft-decision decoding}\label{app:softmbqc}
GKP homodyne residuals supply likelihoods or LLRs in addition to digitized outcomes. Soft-decision decoding retains this information when estimating the logical Pauli class. Known measurement-dependent displacements and discrete corrections are tracked deterministically; uncertainty concerns the residual error, not the recorded correction command. Physical GKP correction and ${\rm C_4}$-level correction or refresh remain quantum operations at their specified circuit locations.
Soft-decision decoding and teleportation-based recursive Bell-state preparation are established ingredients, including the Fibonacci postselection analysis of Aliferis and Preskill and the C4/C6 soft decoder of Goto and Uchikawa~\cite{AliferisPreskill2009Fibonacci,GotoUchikawa2013Soft}. The present contribution does not claim to invent soft decoding; it specifies the measurement-accessible, confidence-ranked selection record at an optical factory--QRL boundary and qualifies the designated data-Bell output under a declared noise protocol. No same-noise benchmark against those earlier decoders is claimed. Section~\ref{sec:interface_contract} defines the correction records, and Sec.~\ref{subsec:postselected_knill} specifies the recursive classical decoder and candidate selection.

\section{Interface between verified resources and online logical operations}\label{sec:interface_contract}

The interface separates what the processor consumes, how matched operations and records realize a logical action, and how well a particular factory supplies the resource. Depth is an operation-dependent execution property, not part of compatibility itself.
We distinguish four mathematical objects throughout: the physical completely positive, trace-preserving (CPTP) preparation and macronode (MN) noise maps $\mathcal N_{\rm prep}$ and $\mathcal N_{\rm MN}$, the outcome-resolved quantum measurement instrument $\{\mathcal E_y\}$, the classical decoder $\mathcal D:y\mapsto p_L(a\mid y)$, and the accepted residual-Pauli error distribution $P_{\rm res,L}(e\mid A_L=1)$. Only under the stated ideal-teleportation/Pauli-diagonal assumptions does the last distribution define an estimated logical Pauli-channel model $\Lambda_{\rm LR}^{(L)}$. Here $A_L=1$ means that the level-$L$ candidate is retained under the declared selection rule; $a_L\in\mathcal A_D$ is its unknown logical Pauli class before the decoder's predicted correction $\hat a_L$, $e_L=\hat a_L\oplus a_L$ is the residual error class, and $E_L$ is the event $e_L\ne0$ on the designated data Bell pair. The acceptance event $A_L=1$ and the failure event $E_L$ are distinct.

\subsection{Interface specification and requirements}
We distinguish the interface specification from a factory's measured performance. The logical-resource interface is
\begin{equation}
\mathsf I_{\rm LR}=
\left(\mathsf R_{\rm log},A_{\rm mode},\mathsf M_{\rm log},
\mathsf F,\mathsf V,\partial_{\rm inj}\right),
\label{eq:logical_resource_interface}
\end{equation}
The interface tuple is code-independent and defines compatibility rather than measured performance. The recursive decoder record $\mathcal R_L$ in Eq.~\eqref{eq:recursive_soft_resource_record} describes one candidate; the separate $\mathcal S_L$ below summarizes its accepted-output \emph{population} under a specified selection and evaluation protocol.
Here $\mathsf R_{\rm log}$ specifies the encoded resource type, $A_{\rm mode}$ assigns code indices to optical space--time addresses, and $\mathsf M_{\rm log}$ specifies the measurement or teleportation operation. The correction-record schema $\mathsf F$ distinguishes a known CV displacement vector $\boldsymbol d_{\rm CV}\in\mathbb R^{2n}$ for $n$ modes, physical-GKP Pauli-frame bits, and the encoded logical Pauli frame (with a Clifford frame when required). MBQC byproducts identify the measurement-dependent origin of such corrections, not an additional independent error. Lattice corrections transferred from the CV record to a Pauli frame are composed once. The schema $\mathsf V$ contains verification observations and decoder-reliability information; $\partial_{\rm inj}$ marks the preparation/online boundary. Compatible factories may differ in error, acceptance, and preparation time.

For the declared numerical scope, a level-$L$ accepted-resource population is characterized by
\begin{equation}
\mathcal S_L=\left(P_L,\,{\rm CI}_L,\,P_{\rm acc}^{(L)},\,R_L^{\rm GKP}\right),
\label{eq:logical_resource_performance_record}
\end{equation}
Throughout the factory characterization, $y_{\rm prep}$ denotes accessible GKP homodyne residuals, verification/Bell outcomes, syndromes, and known addresses, frames, and pre-declared derived features. The index $a\in\{I,X,Y,Z\}$ denotes one of four residual data-Bell Pauli hypotheses prior to the predicted correction. We write $p_L(a\mid y_{\rm prep})$ for their \emph{normalized, decoder-estimated class-score distribution}; it is not automatically a calibrated conditional error probability, nor is it a distribution over unobserved simulator variables. The separate record $y_{\rm post}$ denotes additional online measurement outcomes.
Here $P_L=\Pr(E_L\mid A_L=1)$ is the selected-output logical-error probability, ${\rm CI}_L$ its stated confidence or credible interval, as specified by the statistical method, $P_{\rm acc}^{(L)}=\Pr(A_L=1)$ the acceptance probability at the generated-candidate boundary, and $R_L^{\rm GKP}$ the declared \emph{nominal} supplied-GKP count (Appendix~\ref{app:resource_accounting}). The reported estimates and intervals arise from the independent evaluation counts under their declared statistical conventions. The candidate-specific decoder scores and verification record are components of $\mathcal R_L$, not entries of the population record. The separate estimated Pauli-diagonal model $\Lambda_{\rm LR}^{(L)}$, derived from accepted residual classes under ideal teleportation (Appendix~\ref{app:c4_two_logical_subsystems}), is not a fitted physical noise process; preparation latency and actual delivered-factory cost are not reported as measured fields of $\mathcal S_L$.

The specification is used through six compatibility requirements:
\begin{enumerate}[label=(\roman*),leftmargin=2em,itemsep=2pt,topsep=4pt]
\item \textbf{Resource compatibility.} The accepted resource has the declared encoded type and designated logical subsystem, characterized at $\partial_{\rm inj}$.
\item \textbf{Index/address compatibility.} Physical code index $i$ is mapped to $\alpha_i=(r_i,\tau_i)$, with $r_i$ a spatial optical rail and $\tau_i$ a time-bin index; required pairings and known permutations remain explicit in $A_{\rm mode}$. Rail reuse at different times does not identify the modes because the ordered pair is the address.
\item \textbf{Measurement-role compatibility.} Each participating mode has a declared injection, teleportation, logical-measurement, or routing role, with a defined interpretation of its outcome.
\item \textbf{Frame composability.} Known CV displacements and discrete Pauli/Clifford corrections are composed through $\mathsf F$, with their representation changes tracked without double counting.
\item \textbf{Verification/reliability compatibility.} Syndrome, analog, class-score, and reliability records retain their declared meaning for the consuming decoder or subsequent logical operation.
\item \textbf{Operation compatibility.} The matched quantum operations, measurements, decoding, and address/frame updates together realize the intended encoded action.
\end{enumerate}
These requirements enforce consistency at the resource-to-operation boundary. A noise-inclusive fault-tolerance guarantee requires additional performance characterization. The common element is the Gaussian QRL resource-state family and operation set. A replacement code may require different preparation, routing, buffering, decoding, measurement, or control rules, and its performance record must be characterized separately.

\subsection{Compatible factories and code-dependent implementation}
Candidate factories based on Calderbank--Shor--Steane (CSS) codes, other stabilizer codes, or quantum low-density parity-check (qLDPC) codes must supply and characterize the resource, measurement, address, frame, and reliability elements above for their intended operations~\cite{Raveendran2022QLDPCGKP,Walshe2025ArbitraryCodes}. Compatible factories can supply the same declared logical role while differing in error, acceptance, preparation time, and cost. \emph{Operational equivalence} is stronger: it additionally requires the same conditional logical channel and classical-record distribution at $\partial_{\rm inj}$.

Compatibility and bounded online depth are separate properties. A replacement code with an $O(d)$ logical-measurement protocol may remain compatible with the Gaussian processor while losing the $O(1)$ depth property. Address, measurement, decoding, and control rules must be re-established when the encoding changes.

\subsection{Composability and the logical-state boundary}\label{subsec:interface_composition}
An accepted encoded Bell pair is an ancilla, distinct from the teleported data. An online operation combines encoded-resource injection, QRL measurements, and measurement-conditioned classical processing and corrections. The corresponding quantum-instrument composition and its register definitions are given in Appendix~\ref{app:decoder_details}, Eq.~\eqref{eq:interface_cycle}. This composition specifies the interface; it does not constitute a noisy online logical-gate evaluation. Teleportation and refresh preserve the output register type, not its state. Destructive Bell measurements produce classical outcomes, and the surviving teleported register is tracked separately.

With ideal logical Bell measurements and frame corrections, a Pauli-diagonal encoded Bell resource represents the Choi state of the induced logical teleportation channel; its four logical-Pauli probabilities determine that channel. Noisy injection and subsequent operations require their own channel characterization. Repeated-cycle exclusive-OR (XOR) convolution is used only as a consistency diagnostic when cycles are stationary and memoryless, accepted ancillas are independent, and the residual channel is Pauli-frame diagonal; a complete noise analysis of an online logical operation instead requires composition of the corresponding measurement-conditioned instruments (Appendix~\ref{app:decoder_details}), which is beyond the present Bell-resource numerical evaluation.

\subsection{Logical-state injection}\label{subsec:interface_injection}
Verified encoded resources are coupled to selected macronodes by teleportation-based injection. The physical GKP modes are matched to their corresponding code indices, and the homodyne outcomes supply displacement and Pauli/Clifford updates. Analog residuals and ${\rm C_4}$ syndromes are decoded into logical Bell outcomes and reliability records; residual Pauli uncertainty remains a logical-error contribution after the known address and correction updates.

The index mapping is essential: arbitrary physical controlled-$Z$ (CZ) wiring does not automatically implement a logical CZ.
For each supported construction, a physical QRL instruction acts on a GKP-encoded mode or matched mode pair; transversal use across the ${\rm C_4}$ physical indices is then interpreted through the data/gauge Pauli representatives and the updated address and correction frame. The existence of this ideal logical map does not itself certify a noisy online logical gate. Required pairings and any intra-block permutation must be implemented or tracked explicitly. Table~\ref{tab:address_interface} specifies the mapping $i\leftrightarrow\alpha_i=(r_i,\tau_i)$, which permits rail reuse across time bins without losing the intended transversal pairing.

\begin{table*}[t]
\centering
\small
\renewcommand{\arraystretch}{1.15}
\setlength{\tabcolsep}{7pt}
\begin{tabular*}{\textwidth}{@{\extracolsep{\fill}}p{0.23\textwidth}p{0.71\textwidth}}
\toprule
Interface requirement & Operational definition \\
\midrule
Address assignment & Each physical index $i$ of an encoded block is assigned to a declared QRL space--time address $\alpha_i=(r_i,\tau_i)$. \\
Time-domain reuse & A spatial rail may carry different modes at different time bins, but the full space--time address remains part of the logical-resource record. \\
Transversal pairing & For two-block transversal operations, corresponding physical indices are paired as $A_i\leftrightarrow B_i$, with $i=0,\ldots,n_L-1$ for a block of $n_L$ physical modes ($n_L=4^L$ in the present ${\rm C_4}$ realization). \\
Permutation tracking & If routing or a code automorphism introduces $i\mapsto\pi(i)$, the address record is updated explicitly before the next logical operation. An untracked intra-block permutation is not permitted because it can change the intended physical pairing and logical-subsystem action. \\
Classical interface record & Known address changes and Pauli/Clifford byproducts are tracked classically. Their bookkeeping does not constitute an additional protected QRL/macronode layer, although controller latency and feedforward are accounted for separately. \\
\bottomrule
\end{tabular*}
\caption{Address-preserving interface contract between an offline encoded resource and the online QRL/MBQC processor. The contract preserves the physical-index pairing required by transversal encoded operations while allowing optical rails to be reused across time bins. These routing and bookkeeping requirements support the online operations in Fig.~\ref{concept0}; they are not additional quantum operations.}
\label{tab:address_interface}
\end{table*}

\subsection{Logical-operation constructions and their scope}\label{sec:operations_evidence}
Table~\ref{tab:operation_evidence} gives the physical QRL instruction, encoded ${\rm C_4}$ action, and classical record for each supported construction. Detailed measurement patterns are in Appendix~\ref{app:qrl_logical_operations}, with routing and parallelism assumptions specified by Criterion~\ref{crit:distance_depth}.
Here SUM denotes the continuous-variable controlled-sum gate.

For example, $\operatorname{SWAP}_{23}$ exchanges the ${\rm C_4}$ data ($D$) and gauge ($G$) ports, so the data label follows $D_{\rm in}\mapsto G_{\rm out}$ rather than a fixed physical $D$ port. Recording this permutation identifies the routed output while leaving physical routing and residual logical error to be treated explicitly (Appendices~\ref{sec:concat_c4_stabilizers} and~\ref{app:ds_mapping}).

Non-Clifford resources fit the same interface, but their preparation and distillation are outside the numerical analysis (Sec.~\ref{subsec:universality}).

\begin{table*}[t]
\centering
\small
\renewcommand{\arraystretch}{1.13}
\setlength{\tabcolsep}{5pt}
\begin{tabular*}{\textwidth}{@{\extracolsep{\fill}}p{0.135\textwidth}p{0.285\textwidth}p{0.280\textwidth}p{0.235\textwidth}@{}}
\toprule
\raggedright\textbf{Logical operation} & \raggedright\textbf{QRL instruction and encoded action} & \raggedright\textbf{Classical record and output} & \raggedright\textbf{Construction and evaluated scope} \tabularnewline
\midrule
\raggedright Injection / teleportation &
\raggedright Bell-type homodyne injection of matched GKP modes using an accepted encoded Bell resource. &
\raggedright Addresses, homodyne outcomes, GKP likelihoods, and ${\rm C_4}$ decoding update the surviving data frame and reliability. &
\raggedright Interface and injection construction, Secs.~\ref{subsec:interface_composition}--\ref{subsec:interface_injection}; resource-output evaluation in Sec.~\ref{sec:threshold}. \tabularnewline
\addlinespace[5pt]
\raggedright Logical-subsystem SWAP &
\raggedright A QRL SWAP instruction (four homodyne angles) implements parallel $\operatorname{SWAP}_{23}$ permutations, exchanging the $D$ and $G$ ports. &
\raggedright Track $D_{\rm in}\mapsto G_{\rm out}$ and its frame, rather than retaining a fixed output-port label. &
\raggedright One-layer benchmark: $D_{\rm prot}^{\rm SWAP}=1$, Eq.~\eqref{eq:swap_depth_benchmark}. \tabularnewline
\addlinespace[5pt]
\raggedright Hadamard &
\raggedright A QRL Fourier instruction with the tracked permutation gives $\overline H_D\overline H_G=\operatorname{SWAP}_{23}H^{\otimes4}$ in the stated ${\rm C_4}$ convention. &
\raggedright Compose Fourier/Clifford byproducts and the address permutation to preserve the logical labels. &
\raggedright Fixed-length construction, Appendix~\ref{app:qrl_logical_operations}. \tabularnewline
\addlinespace[5pt]
\raggedright CNOT &
\raggedright Parallel QRL SUM--SWAP on index-matched GKP modes gives CNOTs on $D_A\to D_B$ and $G_A\to G_B$ and exchanges the output registers. &
\raggedright Track block pairing, output exchange, and Pauli/Clifford frames. &
\raggedright Logical map and fixed-length construction, Appendices~\ref{app:ds_mapping} and~\ref{app:qrl_logical_operations}. \tabularnewline
\addlinespace[5pt]
\raggedright Logical Bell measurement &
\raggedright Physical GKP Bell/homodyne measurements and ${\rm C_4}$ parity decoding provide logical Bell outcomes. &
\raggedright Logical outcomes and reliability; update any surviving teleportation output separately. &
\raggedright Measurement/decoding construction, Appendix~\ref{app:qrl_logical_operations}. \tabularnewline
\bottomrule
\end{tabular*}
\caption{Logical actions and classical records at the ${\rm C_4}$/GKP--QRL interface. The ideal maps and tracked corrections follow Appendix~\ref{app:qrl_logical_operations}, with advance resource availability and parallel routing as in Criterion~\ref{crit:distance_depth}. Only SWAP is assigned a numerical depth here. The independent error qualification applies to the designated Bell resource before or after the specified identity-teleportation noise, not to all listed operations.}
\label{tab:operation_evidence}
\end{table*}

\subsection{Parallel resource stacking and online execution}\label{subsec:interface_execution}
The benefit of the stacked arrangement in Fig.~\ref{concept0}(b) is that protection can be increased by enlarging the parallel encoded-resource block while reusing a fixed-length QRL measurement pattern. For the constructions considered here, corresponding physical operations are assigned to matched code-index addresses. Increasing the level replicates those operations across a larger block, rather than appending sequential measurement stages. The interface preserves the pairings, frames, and output addresses needed to interpret these parallel operations as the intended logical action. This makes verified-resource supply and parallel optical capacity explicit scaling parameters while retaining a common Gaussian operation set.

Protected online measurement depth counts sequential QRL/macronode measurement layers acting on encoded data. Classical correction-record processing is excluded from this layer count and included separately in latency. With resources available in advance, concurrent disjoint operations, and address-preserving routing, the native QRL SWAP construction implements the ${\rm C_4}$ permutation $\operatorname{SWAP}_{23}$ in one layer:
\begin{equation}
D_{\rm prot}^{\rm SWAP}=1
\label{eq:swap_depth_benchmark}
\end{equation}

The output record follows $D_{\rm in}\mapsto G_{\rm out}$ so that data identity is not confused with a fixed physical port. This circuit property is evaluated separately from the Bell-resource failure counts. A transversal layer still contains $4^L$ physical operations, and resource preparation and classical decoding can grow with $L$. If parallel capacity is fixed, TDM batches can serialize the layer; shallow protected depth therefore does not alone fix preparation latency, elapsed runtime, or throughput. Appendix~\ref{app:qrl_logical_operations} gives the measurement patterns and full execution criterion; Appendix~\ref{subsec:online_timing} gives the width, timing, and supply-rate relations.

\section{${\rm C_4}$/GKP encoded-resource factory}\label{sec:details}
The factory supplies the quantum resource and classical records required by Sec.~\ref{sec:interface_contract}. Its ${\rm C_4}$/GKP realization is the case study evaluated in Sec.~\ref{sec:threshold}.

\subsection{Preparation and twelve-child verification}\label{sec:circuit_level}
\begin{figure*}[t]
\centering
\includegraphics[width=1\textwidth]{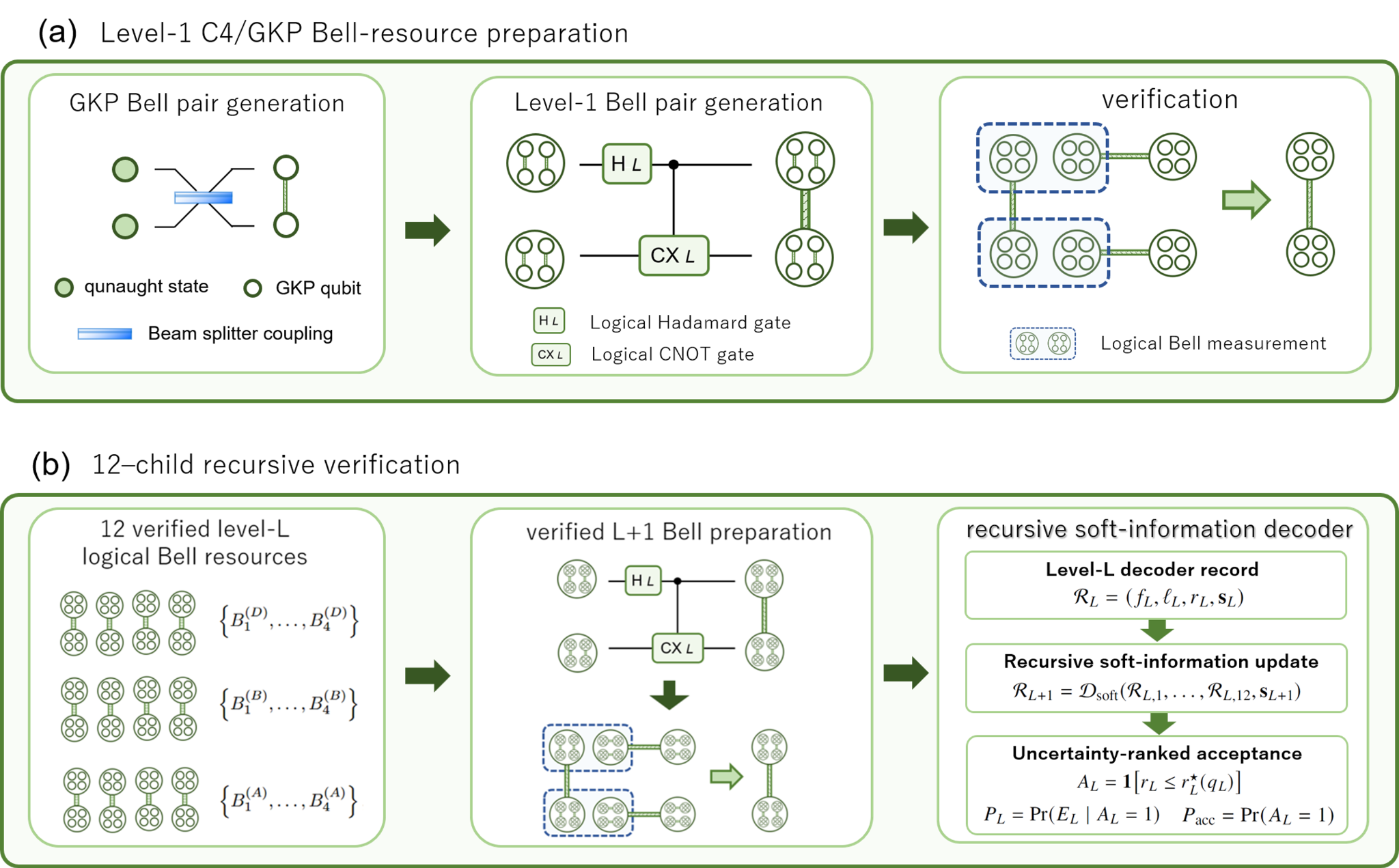}
\caption{${\rm C_4}$/GKP Bell-resource factory. (a) Physical GKP Bell pairs are assembled into a level-1 ${\rm C_4}$/GKP Bell resource by the logical Hadamard/CNOT preparation circuit and verified by logical Bell measurements. (b) Twelve level-$L$ child resources, arranged as one data group and two verification groups, prepare and verify a level-$(L+1)$ candidate. The accept/reprepare decision is schematic: the $12.25$-dB numerical output selection applies a batchwise rank quota rather than a fixed per-candidate risk cutoff (Appendix~\ref{app:selection_policy}). The child records and new verification outcomes determine its decoder class-score record and binary accept/reprepare decision. The tuple in the diagram abbreviates Eqs.~\eqref{eq:recursive_soft_resource_record} and~\eqref{eq:recursive_soft_decoder_map}; the 12 input boxes are level-$L$ resources, whereas a ``class'' is a possible residual logical Pauli error on such a resource; $P_L=\Pr(E_L\mid A_L=1)$ and $P_{\rm acc}=\Pr(A_L=1)$. Non-Clifford-resource preparation is outside this figure.}
\label{fig:factory_recursive}
\end{figure*}

Figure~\ref{fig:factory_recursive}(a) shows level-1 preparation: physical GKP Bell pairs, generated from elementary GKP and qunaught resources (qunaughts are noncomputational GKP grid states), are assembled into a ${\rm C_4}$/GKP block-pair resource by logical Hadamard/CNOT operations and verified by logical Bell measurements. We call a completed level-$L$ encoded Bell-resource candidate, together with its known frame and measurement-derived reliability record, a \emph{child} when it serves as one input to the construction of a level-$(L+1)$ \emph{parent} candidate. A child is not an individual physical ${\rm C_4}$ qubit, and child ``classes'' must not be confused with the logical Pauli-error classes $\mathcal A_D$ defined in Sec.~\ref{sec:model}. At higher levels, a candidate consumes twelve child resources ($N_{\rm child}=12$) in one data group and two verification groups,
\begin{align}
&\{B^{(D)}_1,\ldots,B^{(D)}_4\},\qquad
\{B^{(A)}_1,\ldots,B^{(A)}_4\},\nonumber\\
&\{B^{(B)}_1,\ldots,B^{(B)}_4\}.
\end{align}

In this notation, $B_j^{(g)}$ is the level-$L$ Bell resource occupying slot $j\in\{1,2,3,4\}$ of group $g\in\{D,A,B\}$. The three groups name one data-bearing preparation group ($g=D$) and two verification groups ($g=A,B$); these are circuit roles, not three code types or three error classes. The group superscript $(D)$ is a preparation role and should not be confused with the protected data-qubit subsystem $D$. Each group's four children are ordered by their assigned circuit slots; the $12$ records enter the parent decoder with those slot identities preserved. A \emph{root} is the completed top-level candidate whose designated $D$ Bell output is finally evaluated. Each child supplies its optical state, tracked frame, and reliability record. The fixed teleportation-and-verification network combines these records with new homodyne and syndrome outcomes to form the parent class-score distribution. Exact parity relations, child-slot topology, and address/frame transformations are retained; the decoder and binary acceptance rule are defined in Sec.~\ref{subsec:postselected_knill}. Accepted outputs have the same interface resource type at each level. In the numerical implementation, verification outcomes enter the decoder without an additional syndrome-only hard-rejection gate. Confidence selection acts on completed level-wise outputs, not recursively on every internal child (Appendix~\ref{app:selection_policy}); its nominal cost penalty is specified separately in Appendix~\ref{app:resource_accounting}.

\subsection{Optical implementation}
The layout combines a continuously generated QRL cluster, offline GKP Bell-resource modules, and ${\rm C_4}$ encoding/verification modules. The latter use squeezed-ancilla-assisted quantum-nondemolition (QND) / SUM operations, homodyne detection, and classical feedforward~\cite{Fukui2023NoisyGKPOptical,Yoshikawa2008QNDSumGate,Shiozawa2018QNDRealtime}. In the infinite-ancilla-squeezing and unit-gain limit, the physical operation is
\begin{equation}
U_{\rm SUM}=\exp(-i\hat q_C\hat p_T),
\end{equation}

which implements GKP controlled-$X$ (CX, or CNOT) up to the stated Fourier and frame conventions. Finite ancillary squeezing contributes to the preparation channel; additional macronode noise is applied separately after resource preparation. Source mechanisms, component context, and the noisy physical gate are described in Appendix~\ref{sec:prelim}. The numerical resource metric counts supplied GKP qubits, not a sum of detector, switch, gate, storage, and controller costs.

\subsection{Universality via magic-state injection}\label{subsec:universality}\label{subsec:universality_scope}
The evaluated factory prepares Clifford-type Bell resources. A non-Clifford operation can be supplied by an encoded magic state prepared offline and consumed through logical Bell measurement and frame update~\cite{BravyiKitaev2005Magic,BravyiHaah2012Magic}. The interface accommodates this route, although magic-state preparation and distillation costs are outside the reported numerical resource counts. Optical non-Gaussian implementation routes are summarized in Appendix~\ref{app:optical_resource_options}.

\subsection{Factory-side soft-information decoding and resource selection}\label{subsec:postselected_knill}
Knill-type circuits provide verification measurements~\cite{Knill2005,AGP2008Postselection}. Here $q_L\in(0,1]$ is the prescribed fraction of generated level-$L$ candidates retained by the declared selection policy. Here, verification means assessing each candidate Bell resource from its observed syndromes and analog GKP measurements. For $q_L<1$, we retain the candidates with the lowest decoder-estimated risk; for $q_L=1$, all generated candidates are retained. No additional pass/fail test requiring every syndrome to vanish is imposed.
The fitted decoder assigns four class scores from measurement-accessible factory records. In the frozen-tree inference path, multiclass scores are combined into the designated Bell-Pauli distribution and the known Pauli frame is applied deterministically. \emph{Selection-policy calibration} means choosing the decoder/risk configuration and retained fractions on non-test data and fixing them before independent evaluation. The resulting $r_L=1-\max_a p_L(a\mid y)$ is a candidate-ranking score; no additional post-hoc probability-remapping step is used in this specified construction. Appendix~\ref{app:decoder_policy_calibration} specifies the classical procedure; the distinct physical MN noise model is given in Appendix~\ref{app:qrl_insertion_details}. Appendix Fig.~\ref{Evaluation} summarizes the decoder and the separation between calibration and independent evaluation.

\subsubsection{Recursive record and logical output}\label{subsec:recursive_soft_verification}
A level-$L$ resource carries
\begin{equation}
 \mathcal R_L=
 \left(
 p_L(a\mid y),\;f_L^{\rm known},\;r_L,\;\bm s_L,\;A_{{\rm mode},L}
 \right),
\label{eq:recursive_soft_resource_record}
\end{equation}

Here $p_L(a\mid y)$ is the normalized decoder class-score distribution for the accessible record $y=y_{\rm prep}$ defined in Sec.~\ref{sec:interface_contract}, and $f_L^{\rm known}$ is the encoded Pauli frame after lower-level deterministic updates. The candidate-ranking score $r_L=1-\max_a p_L(a\mid y)$ is smaller for more confidently classified candidates. The syndrome record $\bm s_L$ and mode-address record $A_{{\rm mode},L}$ complete the tuple.
Equation~\eqref{eq:logical_resource_performance_record} summarizes independent population-level error and acceptance statistics, while $\mathcal R_L$ is updated for each candidate. Candidate-level class scores do not themselves yield a Clopper--Pearson bound on the selected population. Twelve child records and new verification outcomes produce
\begin{equation}
 \mathcal R_{L+1}
 =
 \mathcal D^{\rm soft}_{L+1}
 \left(
 \mathcal R_{L,1},\ldots,\mathcal R_{L,12},
 y_{L+1}^{\rm new}
 \right).
\label{eq:recursive_soft_decoder_map}
\end{equation}
Here $\mathcal R_{L,j}$ is the complete record of the child in ordered slot $j$, and $y_{L+1}^{\rm new}$ consists of newly acquired parent-level verification syndromes $\bm s_{L+1}$, measured analog/Bell outcomes, and specified slot/address updates. The child Pauli-class labels are latent hypotheses summarized in these records, not measured child identities; no new parent-level outcome is silently omitted from the map's argument.

${\rm C_4}$ parity constraints, Bell-stabilizer quotienting, child-slot permutations, and deterministic frame composition are handled exactly. The map $\mathcal D_{L+1}^{\rm soft}$ updates classical records; a fitted predictor estimates the remaining class scores from measurement-accessible information. It is not a quantum channel. For the independent level-2 confirmation it is a fixed probabilistic tree ensemble; training, selection-policy calibration, and confirmation use disjoint data.
At a parent node, the specified recursive decoder consumes twelve ordered child records, newly measured verification bits and analog likelihoods, and known slot permutations. It (i) propagates allowed C4 parity/Bell-stabilizer constraints and known frames exactly; (ii) forms or updates the compatible error-class likelihood sums (log-sum-exp at the physical/C4 parity stage); (iii) supplies the resulting observable features to the frozen side-specific multiclass predictors, whose outputs are mapped to four Bell-Pauli class scores as in Eq.~\eqref{eq:frozen_decoder_class_scores}; (iv) applies the known frame and normalizes $p_{L+1}(a\mid y)$, recording $r_{L+1}=1-\max_a p_{L+1}(a\mid y)$; and (v) applies the predeclared candidate-ranking rule. These steps define the components of $\mathcal D^{\rm soft}_{L+1}$ without assuming statistical independence among the twelve child errors. Hidden physical-error labels are not decoder inputs. Appendix~\ref{app:decoder_details} specifies the map and auxiliary diagnostics; physical-to-logical likelihood rules are given in Appendix~\ref{sec:physical_to_c4_llr}.

The residual error is $e_L=\hat a_L\oplus a_L$ after the final deterministic frame update, where $a_L$ is the true logical Pauli class and $\hat a_L$ is the decoder's correction. A logical failure is $e_L\ne0$ on the designated data Bell subsystem. Known corrections are not treated as unknown faults; the full residual-channel definition is given in Appendix~\ref{app:c4_two_logical_subsystems}.

\begin{figure*}[!t]
\centering
\includegraphics[width=\textwidth]{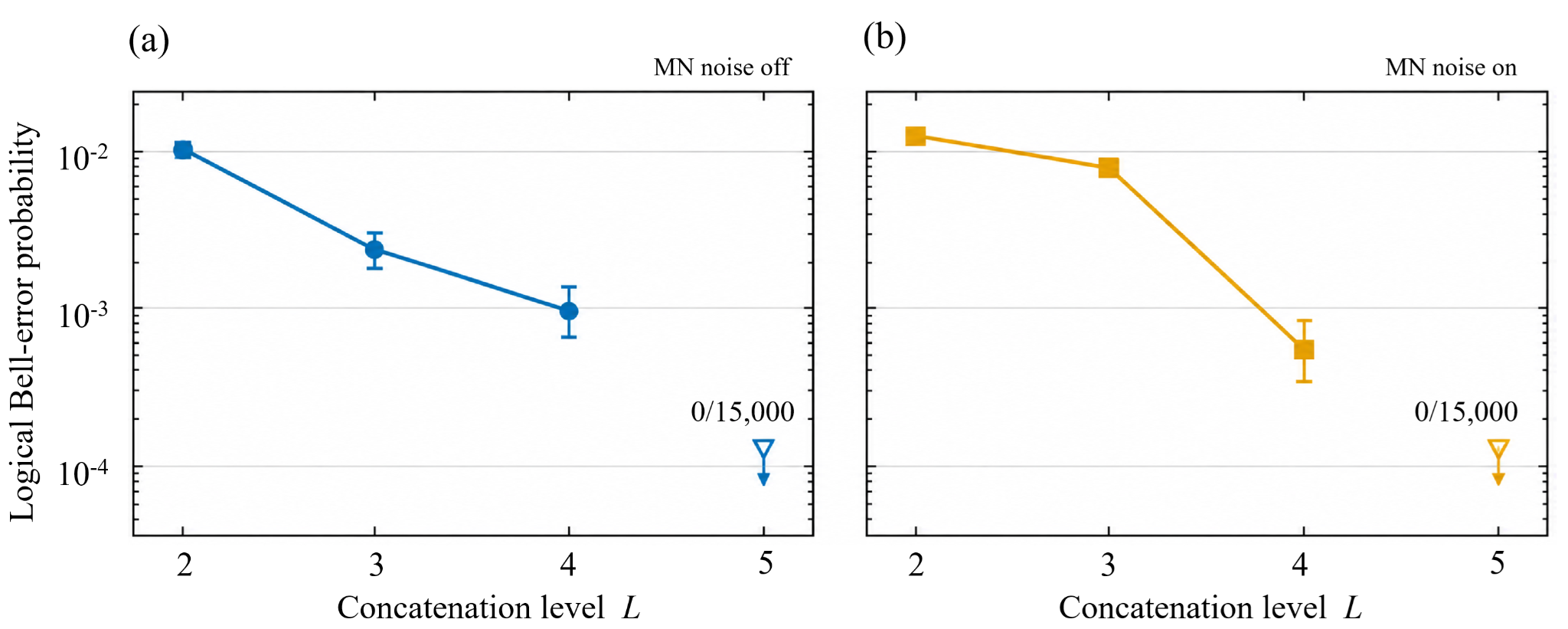}
\caption{Finite-level logical-error contraction at $12.25\,\mathrm{dB}$ for $(q_2,q_3,q_4,q_5)=(1,0.70,0.96,1)$. Confidence selection is applied within each replicate; $q_5=1$ retains all level-5 candidates. The $L=2$--$4$ estimates have $95\%$ Jeffreys intervals; the $L=5$ downward markers are one-sided $95\%$ Jeffreys limits from $0/15{,}000$ accepted samples in each condition. The Jeffreys-posterior ordering probabilities for $P_5<P_4<P_3<P_2$ are $0.99998$ (MN off) and $0.99981$ (MN on). These selected-output results support the stated finite-level ordering under the statistical model of Appendix~\ref{app:statistics}, not a $10^{-5}$ tail.}
\label{fig:fresh_recursive_contraction}
\end{figure*}

\begin{figure*}[!t]
\centering
\includegraphics[width=\textwidth]{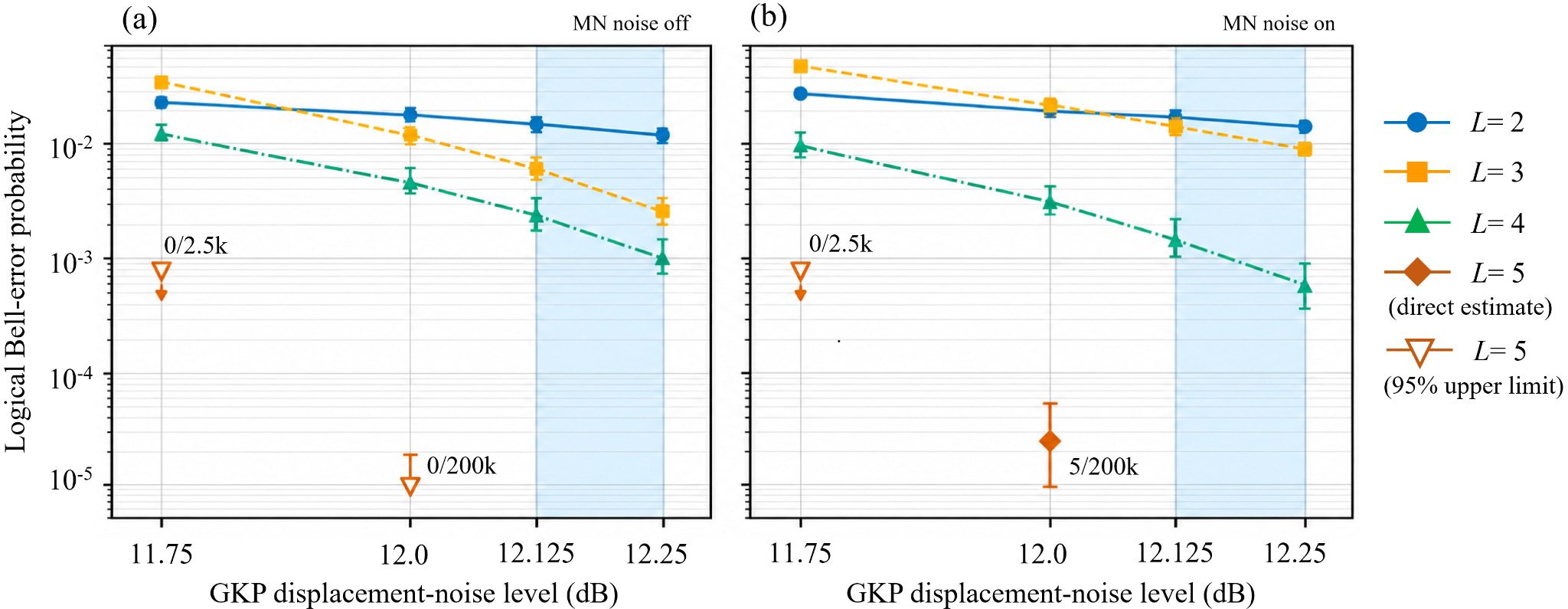}
\caption{Logical Bell-error probability versus GKP displacement-noise level for the fixed schedule. The retention fractions denote the declared numerical rank-selection schedule, not a fixed streaming confidence threshold. Panels (a) and (b) show MN noise off and on. $L=2$--$4$ estimates have $95\%$ Jeffreys intervals. The sparse $L=5$ support points are restricted to $11.75$ and $12.00$ dB: inverted triangles denote zero-event upper limits; the diamond is $5/200{,}000=2.5\times10^{-5}$. No $L=5$ trend line is inferred. Shading marks the evaluated-grid transition of the posterior ordering diagnostic, not an interpolation of these curves. The independent $12.25$-dB ordering sample remains in Fig.~\ref{fig:fresh_recursive_contraction}; support-point counts and intervals are detailed in Appendix~\ref{app:curve_support}.}
\label{fig:recursive_contraction_results}
\end{figure*}

\subsubsection{Binary selection using decoder-estimated risk}\label{subsec:sp_knill_soft}
At $12.25$ dB, $q_L$ is the prescribed fraction retained from generated candidates within each replicate, level, and MN condition. A replicate is one separately ranked candidate batch for a fixed level and noise condition, and $M$ denotes the number of generated candidates in that batch ($M=500$ for the stated level-2--4 ordering batches). Candidates are ranked by increasing risk; $\max\{1,\lceil Mq_L\rceil\}$ of $M$ candidates are retained. At the reported batch sizes this gives discard fraction $1-q_L$. This quota fixes no numerical risk cutoff.

For the declared candidate-selection rule, the selected-output error probability and the candidate-level retention probability are
\begin{equation}
P_L(q_L)=\Pr(E_L\mid A_L=1),
\qquad
P_{\rm acc}^{(L)}(q_L)=\Pr(A_L=1).
\label{eq:soft_information_conditional_error}
\end{equation}

Acceptance is binary; the fitted decoders, retained fractions, and tie rules are fixed before evaluation. At $q_L=1$, all generated candidates in the soft-verification mode are retained. Resource errors and experimental source losses remain possible. The reported acceptance is defined relative to the generated-candidate boundary rather than as a source-to-processor delivery probability. Appendices~\ref{app:selection_policy} and \ref{app:qualification_diagnostics} specify the numerical selection rules; Appendix~\ref{app:resource_accounting} distinguishes nominal cost from complete expected consumption.

\section{Online injection and decoding interface}\label{sec:online_decoder_interface}
Section~\ref{sec:details} characterizes \emph{preparation}: the offline factory creates Bell candidates, propagates their analog and syndrome records, performs final resource GKP correction, and retains candidates according to the declared policy. The consuming processor instead acts on the injected data and produces \emph{new} measurement outcomes. Its decoder must update the logical frame in the presence of those outcomes; the present numerical results do not constitute a separately simulated, end-to-end noisy online-gate qualification.

\subsection{Accessible online record and logical correction}
Let $y_{\rm prep}$ denote the frozen record delivered with the accepted $D$-Bell resource. After injection, $y_{\rm post}$ contains newly available Bell/homodyne quadratures and modular residuals, measured syndrome outcomes where prescribed, and the known address, gauge-role, and Pauli/Clifford-frame changes. An online decoder may evaluate
\begin{equation}
p_{\rm on}(a\mid y_{\rm prep},y_{\rm post},\mathsf F_{\rm known})
\label{eq:online_decoder_conditional_record}
\end{equation}
for the residual data-Pauli class $a$. The factory record is conditioned on rather than counted as a new independent measurement. A measured GKP modular residual enters as an even/odd wrapped-Gaussian likelihood, not as an idealized hidden displacement or error label. The C4 parity constraints then map those physical likelihoods and the Bell outcomes to a data-qubit four-class score distribution, followed by deterministic frame/address composition and a declared correction or logical-outcome interpretation.

\subsection{Operational decoding sequence and scope}
For clarity, a possible measurement-accessible online processing contract is:
\begin{enumerate}[label=(\roman*),leftmargin=2em,itemsep=2pt,topsep=2pt]
\item Acquire the new Bell/homodyne outcomes and compute modular residual likelihoods under the specified physical noise convention.
\item Condition on the delivered factory record without reusing its analog samples as statistically fresh evidence; propagate C4 parity constraints and any tracked data--gauge role permutation.
\item Infer the four-class residual data-Pauli estimate using the decoder assigned to this measurement instrument and update the known Pauli/Clifford frame.
\item Report the logical outcome and its reliability; a physical logical correction/refresh is performed only where the circuit explicitly calls for it.
\end{enumerate}
This is an \emph{interface specification} for applying analog information after injection. Its online conditional model, calibration and noise-inclusive gate failure rate are not established by the frozen Bell-resource decoder or its independent $15.6$-dB samples.

The ideal identity/Bell-type homodyne injection, Fourier--permutation, index-matched SUM--SWAP, and GKP Bell-measurement instructions are mapped to their ${\rm C_4}$ actions and frame updates in Table~\ref{tab:operation_evidence}. These instructions use the new online measurement record described above; their QRL patterns appear in Appendix~\ref{app:qrl_logical_operations}.

\section{Numerical characterization of the supplied verified resource}\label{sec:threshold}
We generate noisy Bell-resource candidates, apply final GKP correction and frozen classical decoding, select the prescribed outputs, and count residual Pauli failures on the designated data subsystem. Preparation includes GKP displacement noise and finite-ancilla-squeezing QND/SUM noise. The MN-noise-off and MN-noise-on conditions differ only by a prescribed single-step additive-displacement channel applied after resource preparation. The off condition adds no additional displacement noise, whereas the on condition adds independent Gaussian displacements with variance $4\,10^{-2/10}\sigma_{\rm GKP}^{2}(s_{\rm GKP})$ in each relevant quadrature; the factor $10^{-2/10}$ corresponds to a $2\,\mathrm{dB}$ increase in the reference GKP noise-quality parameter. Both conditions retain preparation noise and final GKP correction. This effective noise step is distinct from a device-calibrated macronode gate-noise model (Appendix~\ref{sec:mbqc_insertion_noise}). For the headline physical-coordinate test, the order is candidate preparation and recording, the additional displacement draw (MN on only), final physical GKP correction, and fixed logical decoding, followed by the declared candidate selection and the designated-output failure test. Preparation-stage record updates are distinct from this final decoder pass. 

\subsection{Finite-level contraction criterion near 12 dB}
For the fixed retained-fraction schedule $\FinalRecursiveSchedule$, per-replicate selection at $12.25$ dB (Appendix~\ref{app:selection_policy}) gives selected-sample estimates exhibiting
\begin{equation}
P_{5}^{(m)}
< P_{4}^{(m)}
< P_{3}^{(m)}
< P_{2}^{(m)}.
\label{eq:recursive_l2_l5_ordering}
\end{equation}

separately for MN off and on, with $m\in\{\mathrm{off},\mathrm{on}\}$ denoting the prescribed additional-noise condition and $P_L^{(m)}$ the corresponding selected-output failure probability. Here PS abbreviates postselection, and ${\rm PS,prep}$ designates the selected preparation-only Bell-resource error, while ${\rm MN,post}$ denotes the corresponding error with the specified extra MN displacement and final GKP correction:
\begin{equation}
P_L^{(m)}\equiv
\begin{cases}
P_L^{\rm PS,prep}, & m=\mathrm{off},\\
P_L^{\rm MN,post}, & m=\mathrm{on},
\end{cases}
\label{eq:condition_specific_ordering_metric}
\end{equation}

The generated-candidate retention fractions are $100\%$, $70\%$, $96\%$, and $100\%$ at $L=2,3,4,5$, respectively. The Jeffreys-posterior probabilities of the strict ordering are $0.99998$ and $0.99981$, respectively (Appendix~\ref{app:statistics}). Figure~\ref{fig:fresh_recursive_contraction} shows the level dependence. At $L=5$, no failures are observed in $15{,}000$ accepted samples for either condition, giving a one-sided $95\%$ Jeffreys limit of $1.28\times10^{-4}$; this bound neither establishes zero error nor constitutes a $10^{-5}$ qualification.
Because decoder configurations and retained candidate populations can differ between the two MN conditions, their selected-output error rates do not provide a controlled comparison of the added physical noise under an identical decoder and selection rule.

The adjacent $12.125$-dB point does not pass the ordering criterion for both conditions; the evaluated-grid transition is
\begin{equation}
12.125~\mathrm{dB}<s_{\rm rec,grid}^{\star}\leq12.25~\mathrm{dB}.
\label{eq:recursive_grid_transition_interval}
\end{equation}

Accordingly, the result is a fixed-schedule, finite-level grid criterion rather than a continuous physical threshold or conventional physical--logical pseudothreshold. The native level-1 source uses a different preparation map and is not part of this recursive ordering test. Figure~\ref{fig:recursive_contraction_results} gives the displacement-noise dependence and separate sparse $L=5$ support points; their statistical role is detailed in Appendix~\ref{app:curve_support}.

\subsection{\texorpdfstring{Independent $10^{-5}$ qualification at 15.6 dB}{Independent 10-to-the-minus-five qualification at 15.6 dB}}\label{sec:direct_target_qualification}
A diagnostic scan selected $15.6$ dB for a separate fixed-sample level-2 confirmation with $q=1$. The diagnostic samples were not pooled into this test, and no outcome-dependent sample extension or decoder reselection was used (Appendix~\ref{app:qualification_diagnostics}). Under the fixed-sample binomial model of Appendix~\ref{app:statistics}, with $10^6$ samples per condition, the observed failures and one-sided $95\%$ Clopper--Pearson upper bounds are
\begin{align}
\text{MN noise off}:&\quad 3/10^6,\quad U_{0.95}=7.7536\times10^{-6},\\
\text{MN noise on}:&\quad 4/10^6,\quad U_{0.95}=9.1535\times10^{-6}.
\end{align}

Under this model, each specified condition separately meets the $10^{-5}$ target; these are pointwise statements rather than a joint familywise $95\%$ guarantee. Table~\ref{tab:qualified_operating_points} places this result alongside the other evaluated high-squeezing operating points, which retain their stated Jeffreys convention. The level-5 ordering and support samples are separate from this confirmation.

\begin{table*}[t]
\centering
\small
\setlength{\tabcolsep}{4.5pt}
\begin{tabular*}{\textwidth}{@{\extracolsep{\fill}}ccccccc}
\toprule
$s_{\rm GKP}$ (dB) & $(L,q_L)$ & interval & $k_{\rm off}/n_{\rm off}$ & $k_{\rm on}/n_{\rm on}$ & status & $R_{\rm GKP}$ \\
\midrule
13.0 & $(3,0.70)$ & Jeffreys 95\% & $0/504000$ & $24/504000$ & one condition above target & 4937 \\
14.0 & $(3,0.95)$ & Jeffreys 95\% & $0/199500$ & $0/199500$ & below target under stated convention & 3638 \\
15.0 & $(2,1)$ & Jeffreys 95\% & $26/10^6$ & $26/10^6$ & both above target & 288 \\
15.6 & $(2,1)$ & Clopper--Pearson 95\% & $3/10^6$ & $4/10^6$ & below target under binomial model & 288 \\
16.0 & $(2,1)$ & Jeffreys 95\% & $0/200000$ & $0/200000$ & below target under stated convention & 288 \\
\bottomrule
\end{tabular*}
\caption{Evaluated operating points for one accepted designated data-subsystem Bell resource. Off/on refer to the two additional-MN-noise conditions. Under the fixed-sample binomial model, the $15.6$-dB Clopper--Pearson upper bound is below $10^{-5}$ separately for each condition; other rows retain their specified one-sided Jeffreys convention. $R_{\rm GKP}$ is the nominal supplied-GKP count, not a measured complete retry cost. Statistical assumptions and accounting are defined in Appendices~\ref{app:statistics} and~\ref{app:resource_accounting}.}
\label{tab:qualified_operating_points}
\end{table*}

\subsection{Input-quality versus factory-resource tradeoff}\label{sec:resource_accounting}\label{sec:rhg_scope}
The $14$-dB level-3 point with $q_3=0.95$ meets the target under its Jeffreys convention, whereas the $15.6$-dB level-2 point is independently qualified by the exact-binomial criterion. Their nominal supplied-GKP counts are $3637.9$ and $288$ per accepted designated Bell pair, respectively, giving a within-factory ratio of $12.63$. The higher-quality input thus permits a lower concatenation level at the target error under the stated nominal accounting convention.

Quality-dependent source costs and complete factory consumption are not included; Appendix~\ref{app:resource_accounting} specifies the cost recurrence and break-even condition. Supplementary tail diagnostics and the separate ${\rm C_4}$/Raussendorf--Harrington--Goyal (RHG) comparison boundaries appear in Appendices~\ref{app:qualification_diagnostics} and~\ref{app:rhg_parameters}.

\section{Discussion and conclusion}\label{sec:disc}
The proposed factory--processor interface links resource type, mode addresses, measurement roles, frames, and reliability records to an intended logical action. Its ${\rm C_4}$/GKP realization connects verified Bell resources to injection, SWAP, Hadamard, CNOT, and logical Bell measurement on a common Gaussian QRL processor.
The supported Clifford constructions exploit \emph{transversal logical operations} where applicable: corresponding physical QRL measurement patterns act in parallel on matched code-index addresses, while code automorphisms and measurement byproducts are explicitly tracked. The address-tracked one-layer logical-subsystem SWAP provides a permutation benchmark; the index-matched SUM component of the QRL SUM--SWAP pattern supplies the transversal logical CNOT construction. With encoded resources prepared in advance and sufficient address-preserving routing, stronger protection enlarges the parallel encoded-resource stack without extending the protected online measurement sequence of these supported constructions. Preparation, finite parallel capacity, and classical decoding remain level-dependent costs; hardware serialization can increase elapsed runtime despite the protected-depth property. 
The selected-output numerical case study shows finite-level error suppression and separately qualifies a designated level-2 Bell resource below $10^{-5}$ under the two stated noise conditions. The high-squeezing operating points illustrate an input-quality versus nominal factory-cost tradeoff. The resource results complement the logical-operation constructions; an all-gate error certificate would require additional noisy-operation analysis.

Compatible changes to preparation, verification, decoding, or encoding can be assessed at the same factory--processor boundary, with code-dependent address, measurement, and control rules updated and each factory's performance characterized independently. Two-subsystem outputs, non-Clifford factories, storage loss, and algorithm-level timing remain beyond the present evaluation. Together, the interface specification and the ${\rm C_4}$/GKP case study establish a concrete architectural framework for assessing verified-resource preparation and online logical execution on a common Gaussian optical processing platform. The factory--processor boundary makes resource quality, code-dependent control, and the conditions for shallow online processing explicit and testable.

\begin{acknowledgments}

This work was supported by JST Moonshot R\&D (Grant No. JPMJMS256I). This paper is based on
results obtained from a project, JPNP20017, subsidized
by the New Energy and Industrial Technology Development Organization (NEDO).

The authors disclose financial competing interests:
Kosuke Fukui, Takaya Matsuura, Yohji Chin, Shohei Kiryu, and Takaya Hoshi are employees of OptQC
Corp., and Kan Takase and Warit Asavanant are directors of, and hold equity in, OptQC Corp. In addition,
OptQC Corp. has filed a patent application related to
the work described in this manuscript.

\end{acknowledgments}

\noindent\textbf{Data Availability.---} The data and software supporting the findings of this study cannot be made publicly available because they contain commercially sensitive information. They are available from the corresponding author upon reasonable request.

\appendix
\section{GKP and optical-operation conventions}\label{sec:prelim}
This appendix fixes the conventions and minimal operations used by the main text. We use quadratures satisfying $[\hat q,\hat p]=i$ and vacuum variance $1/2$. GKP-state imperfections and gate noise are represented, unless stated otherwise, by additive Gaussian displacement noise.

\subsection{GKP qubits}
\label{app:gkp_operations}

We use the standard square-lattice GKP convention~\cite{GKP2001}. In the \(q\)-quadrature representation, the ideal logical basis states are Dirac combs,
\begin{align}
 \ket{0_{\GKP}}
 &\propto
 \sum_{s\in\mathbb{Z}}
 \ket{q=2s\sqrt{\pi}},
 \label{eq:appA_gkp_zero_q}\\
 \ket{1_{\GKP}}
 &\propto
 \sum_{s\in\mathbb{Z}}
 \ket{q=(2s+1)\sqrt{\pi}}.
 \label{eq:appA_gkp_one_q}
\end{align}
The conjugate-basis codewords are obtained by the corresponding combs in the \(p\) quadrature. A displacement by \(\sqrt{\pi}\) in \(q\) changes the logical \(X\) frame, while \(\exp(i\sqrt{\pi}\hat q)\) distinguishes the two logical \(Z\)-eigenstates.

The ideal comb states have infinite energy. Finite-energy approximations can be described by Gaussian-broadened peaks and a finite envelope, but the reported Monte Carlo calculations do not propagate these approximate pure-state wavefunctions. Instead, for error propagation they represent each noisy GKP qubit as an ideal encoded state followed by an effective independent Gaussian-displacement channel,
\begin{equation}
 \rho_{\sigma}
 =
 \mathcal{N}_{\sigma}
 \left(\rho_{\rm ideal}\right).
 \label{eq:appA_gkp_finite_energy_channel}
\end{equation}
The noise channel is
\begin{align}
 \mathcal{N}_{\sigma}(\rho)
 &=
 \int d\xi_q\,d\xi_p\,
 P_{\sigma}(\xi_q,\xi_p)
 D(\xi_q,\xi_p)
 \rho
 D^{\dagger}(\xi_q,\xi_p),
 \label{eq:appA_gkp_random_displacement_channel}\\
 P_{\sigma}(\xi_q,\xi_p)
 &=
 \frac{1}{2\pi\sigma_{\GKP}^2}
 \exp\!\left[
 -\frac{\xi_q^2+\xi_p^2}
 {2\sigma_{\GKP}^2}
 \right].
 \label{eq:appA_gkp_gaussian_noise_distribution}
\end{align}
Here
\begin{equation}
 D(\xi_q,\xi_p)
 =
 \exp(-i\xi_q\hat p)
 \exp(i\xi_p\hat q)
 \label{eq:appA_gkp_displacement_operator}
\end{equation}
is the phase-space displacement operator. Equivalently, the measured quadratures are modeled as
\begin{align}
 q &\mapsto q+\xi_q,
 &
 p &\mapsto p+\xi_p,
 \label{eq:appA_gkp_gaussian_shift_model}\\
 \xi_q,\xi_p
 &\sim
 \mathcal{N}(0,\sigma_{\GKP}^2).
 \nonumber
\end{align}
The shorthand $\sigma_{\GKP}^2$ denotes $\sigma_{\GKP}^2(s_{\rm GKP})$ in Sec.~\ref{sec:gkp_interface}; this effective input-noise parameter is distinct from optical cluster-source squeezing. The same variance convention is used for the GKP-displacement, finite-squeezing gate-noise, and additional MN-noise terms.

GKP correction reduces measured quadratures modulo the lattice spacing. For a scalar Gaussian shift of variance $v$, the odd-bin error probability is $P_{\rm odd}(v)\equiv P_{\rm err}(\sqrt v)$, using the single definition in Eq.~\eqref{eq:gkp_digitized_error_main}.

When the analog residual \(\xi\) is retained, the corresponding GKP log-likelihood ratio is
\begin{align}
 \ell_{\GKP}(\xi;v)
 &=
 \log
 \frac{
 \displaystyle
 \sum_{m\in 2\mathbb{Z}}
 \exp\!\left[
 -\frac{(\xi-m\sqrt{\pi})^2}{2v}
 \right]
 }{
 \displaystyle
 \sum_{m\in 2\mathbb{Z}+1}
 \exp\!\left[
 -\frac{(\xi-m\sqrt{\pi})^2}{2v}
 \right]
 }.
 \label{eq:appA_gkp_llr}
\end{align}
The sign of \(\ell_{\GKP}\) gives the most likely GKP Pauli frame, and its magnitude gives the confidence. These analog likelihoods are used by the residual logical Pauli error decoder and by the decoder-risk postselection rule. Thus, even when the final logical variable is binary, the decoder uses continuous GKP information before the modulo reduction.

\subsection{Physical QND/SUM gate and logical GKP CX}
\label{app:qnd_cnot_model}

The ${\rm C_4}$ factory uses physical Gaussian entangling operations that implement GKP CX gates. In the optical model adopted here, the physical gate is the noisy CV QND/SUM gate considered in Ref.~\cite{Fukui2023NoisyGKPOptical}, implemented by two input modes $C,T$ and two ancillary squeezed-vacuum modes $A,B$. With beam-splitter parameter $R$ and ancillary squeezing parameters $r_A,r_B$, the quadratures transform as
\begin{align}
\hat q_C'&=\hat q_C-
\sqrt{\frac{1-R}{1+R}}\,\hat q_A e^{-r_A},
\nonumber\\
\hat p_C'&=\hat p_C-
\frac{1-R}{\sqrt R}\hat p_T
+
\sqrt{\frac{R(1-R)}{1+R}}\,\hat p_B e^{-r_B},
\nonumber\\
\hat q_T'&=
\frac{1-R}{\sqrt R}\hat q_C+\hat q_T
+
\sqrt{\frac{R(1-R)}{1+R}}\,\hat q_A e^{-r_A},
\nonumber\\
\hat p_T'&=\hat p_T+
\sqrt{\frac{1-R}{1+R}}\,\hat p_B e^{-r_B}.
\label{eq:noisy_qnd_heisenberg}
\end{align}
Choosing $(1-R)/\sqrt R=1$ and taking $r_A,r_B\to\infty$ gives the ideal controlled-SUM operation, which implements a logical GKP CX up to local Fourier-frame conventions. For finite squeezing, the ancillary noises increase the GKP displacement variances. If
\begin{equation}
\sigma_A^2=\frac12e^{-2r_A},
\qquad
\sigma_B^2=\frac12e^{-2r_B},
\end{equation}
then the variance update used for preparation-noise effect is
\begin{align}
\sigma_{C,q}^{2}\;&\mapsto\;
\sigma_{C,q}^{2}
+
\frac{1-R}{1+R}\sigma_A^2,
\nonumber\\
\sigma_{C,p}^{2}\;&\mapsto\;
\sigma_{C,p}^{2}
+
\frac{(1-R)^2}{R}\sigma_{T,p}^{2}
+
\frac{R(1-R)}{1+R}\sigma_B^2,
\nonumber\\
\sigma_{T,q}^{2}\;&\mapsto\;
\sigma_{T,q}^{2}
+
\frac{(1-R)^2}{R}\sigma_{C,q}^{2}
+
\frac{R(1-R)}{1+R}\sigma_A^2,
\nonumber\\
\sigma_{T,p}^{2}\;&\mapsto\;
\sigma_{T,p}^{2}
+
\frac{1-R}{1+R}\sigma_B^2 .
\label{eq:noisy_qnd_variances}
\end{align}
This physical QND/SUM noise is part of $\mathcal N_{\rm prep}^{(L)}$, not part of the QRL post-injection channel. The latter is applied only after a logical resource has been prepared and decoded. In a component-level implementation model, each physical QND/SUM gate also consumes ancillary squeezed modes and homodyne/feedforward operations; these quantities are not part of the principal supplied-GKP metric.

\subsection{Optical resource-source options}\label{app:optical_resource_options}
The numerical model specifies effective GKP displacement noise, not a microscopic state-generation route. Possible sources include breeding, photon-counting/catalysis, nonlinear optical or free-electron methods, and traveling-wavepacket or deterministic schemes~\cite{Vasconcelos2010,Weigand2018GridStates,Eaton2019PhotonCatalysis,Tzitrin2020ApproxGKP,Eaton2022PhANTM,Hastrup2022,Fukui2022CrossKerr,Fukui2022Backcasting,Takase2023GaussianBreeding,Dahan2023FreeElectronGKP,Takase2024FlyingGKP,Pizzimenti2024SSCBreeding,Winnel2024DeterministicGKP,Erkilic2026UnifiedGKP}. Relevant components include squeezed sources and waveguide optical parametric amplifiers (OPAs)~\cite{Takeno2007NineDbSqueezing,Vahlbruch2016FifteenDbSqueezing,Kashiwazaki2020SixDbTHzPPLN,Kashiwazaki2023Over8DbWaveguide,Hirota2026TenDbWaveguide,Ha2026TwelveDbWaveguide}, time-resolved non-Gaussian states~\cite{Gerrits2010NumberResolvedPhotonSubtraction,Yukawa2013ThreePhotonSuperposition,Takase2022WaveguideCat,Sonoyama2023TimeGated,Endo2023MultiPhotonSubtractionTelecom,Sonoyama2024MultiPhotonFock,Kawasaki2024BroadbandNonGaussian}, and measurement-induced or OPA-based QND interactions~\cite{Filip2005MeasurementInducedQND,Yoshikawa2008QNDSumGate,Shiozawa2018QNDRealtime,Yanagimoto2023QNDOPA}.

Non-Clifford optical options include cubic-phase, higher-order and nonlinear-quadrature gates, nonlinear squeezing, and adaptive feedforward~\cite{Yukawa2013CubicNonlinearity,Miyata2016AdaptiveCubicGate,Marek2018NonlinearQuadratureGates,Hanamura2024MultimodeCVGates,Konno2021NonlinearSqueezing,Konno2021NonCliffordGKP,Sakaguchi2023NonlinearFeedforward,Yanagimoto2020KerrCubic}. Neither their preparation costs nor source-specific generation costs enter the reported supplied-GKP metric. Analog decoding and concatenation also occur in quantum-repeater settings~\cite{Fukui2021GKPRepeater,Rozpedek2021ConcatenatedRepeater,Rozpedek2023AllPhotonicGKPRepeater,Schmidt2024GKPQuditRepeater,Haussler2025StationaryGKPRepeater,Haussler2026StationaryFlyingGKPRepeater}.

\section{${\rm C_4}$ data--gauge code and logical maps}\label{app:c4_subsystem_rules}
This Appendix specifies the $[[4,1,1,2]]$ data--gauge realization used in the main text and the physical/code operations required to interpret the supplied $D$-Bell pair. The $[[4,2,2]]$ notation continues to denote the same underlying four-qubit stabilizer space.
\subsection{${\rm C_4}$ code and soft concatenation rules}\label{sec:concat_c4_stabilizers}
The underlying $[[4,2,2]]$ ${\rm C_4}$ stabilizer block is used here as a $[[4,1,1,2]]$ subsystem code, with one designated data qubit and one tracked gauge qubit~\cite{Knill2005,AGP2008Postselection}. Its Clifford action under physical-qubit permutations is tracked explicitly rather than inferred from transversality alone~\cite{Sayginel2025C4Clifford}. The stabilizer generators and data/gauge Pauli representatives are fixed in Eq.~\eqref{eq:c4_data_gauge_main}; the same convention is used in the following logical maps.
The gauge group is $\mathcal G=\langle S_X,S_Z,\bar X_G,\bar Z_G\rangle$ (phases implicit). Its center modulo phases is generated by $S_X,S_Z$; the data Pauli algebra commutes with the gauge group. The dressed logical distance is two. This subsystem choice does not assert that the gauge qubit is initialized, protected, or independently qualified as a second delivered logical output.

With the chosen data/gauge representatives, bare transversal Hadamard exchanges the two Pauli algebras,
\begin{align}
H^{\otimes4}:\quad
\bar X_D&\mapsto\bar Z_G, & \bar Z_D&\mapsto\bar X_G,\\
\bar X_G&\mapsto\bar Z_D, & \bar Z_G&\mapsto\bar X_D.
\end{align}
The physical permutation $\operatorname{SWAP}_{23}$ also exchanges $D$ and $G$. Their composition therefore gives the label-preserving logical Hadamard
\begin{equation}
\bar H_D\bar H_G=\operatorname{SWAP}_{23}H^{\otimes4}.
\label{eq:label_preserving_H}
\end{equation}
The same physical permutation $\operatorname{SWAP}_{23}$ implements the logical exchange $D\leftrightarrow G$ under the Pauli representatives above.
In the subsystem interpretation, this exchanges the roles of the data and gauge algebras. It is not a gauge-only operation preserving a fixed protected data algebra; the output address must explicitly redefine which algebra carries the information. Consequently the one-layer permutation is an ideal code-space/address benchmark, not an independently quantified protected SWAP on two arbitrary information qubits. Corresponding physical CNOTs between two ${\rm C_4}$ blocks implement independent logical CNOTs on $D$ and $G$. Whether a gauge-qubit fault contributes to the designated output therefore depends on the exact logical map and output-address convention: $\operatorname{SWAP}_{23}$ exchanges the subsystem labels, whereas the ideal transversal CNOT preserves the $D/G$ partition. Appendix~\ref{app:ds_mapping} gives these mappings explicitly. This corrects the ambiguity that would arise from treating either bare $H^{\otimes4}$ or an untracked physical permutation as label preserving.

The code detects any single-qubit Pauli error but does not by itself correct arbitrary single-qubit errors without additional syndrome and verification structure. This is why the architecture uses ${\rm C_4}$ as a compact verified-resource layer rather than as a standalone memory code.

Under ideal concatenated-code counting, an $L$-level ${\rm C_4}$ block has physical-resource count
\begin{equation}
 n_{\rm C4}(L)=4^L,
\qquad
 d_{\rm C4}(L)=2^L,
\qquad
 n_{\rm C4}(L)=d_{\rm C4}(L)^2 .
\end{equation}
The actual verified-resource cost of the Knill-type recursive factory is larger than this resource count because a level-$(L+1)$ verified resource consumes multiple level-$L$ resources, verification ancillae, measurements, classical decoding, and possibly buffering. The resource discussion therefore distinguishes the nominal supplied-GKP cost from the bare code-block size.

For soft decoding, a measured ${\rm C_4}$ syndrome $\bm s$ defines the compatible physical error sets $\mathcal E_{0}(\bm s)$ and $\mathcal E_1(\bm s)$ for the two logical-error hypotheses. For a fixed $X$- or $Z$-error component, let $\bm e=(e_1,\ldots,e_4)\in\{0,1\}^4$ denote the component-wise physical-error pattern. Then $\mathcal E_b(\bm s)$ consists of patterns compatible with the observed syndrome and inducing designated-data logical parity $b\in\{0,1\}$, with gauge alternatives summed over rather than declared to be additional delivered outputs. These $b$ labels are binary parity hypotheses, not child-group labels or the full four-class Pauli alphabet. The displayed product-likelihood form presumes conditionally factorized incoming binary likelihoods; correlated inputs require joint-pattern weights. The corresponding designated-data logical log odds are obtained by the syndrome-constrained log-sum-exp marginalization in Eq.~\eqref{eq:c4_llr}; the compatibility and conditional-factorization assumptions above apply to that expression.
This same operation is recursively applied in the concatenated factory: physical GKP analog likelihoods are combined with the ${\rm C_4}$ syndrome constraints, lower-level decoder records become inputs to higher-level verification gadgets, and every level outputs the formal recursive record $\mathcal R_L$ defined in Eq.~\eqref{eq:recursive_soft_resource_record}.

The twelve-child construction used in the simulations is therefore interpreted as a recursive soft-information resource map,
\begin{equation}
\left(\{\mathcal R_{L,j}\}_{j=1}^{12},\bm s^{(L+1)}\right)
\longmapsto
\mathcal R_{L+1},
\end{equation}
with the transition constraints determined by the teleportation and verification circuit; the full circuit-indexed transition table is not reproduced here. The present paper reports the decoder-dependent logical-Bell preparation analysis; the physical QND/SUM gate used for each GKP CX is specified above, while the resource discussion uses the supplied-GKP nominal schedule-cost convention defined in Sec.~\ref{sec:resource_accounting}.
\subsection{Likelihood propagation through ${\rm C_4}$ verification}\label{sec:c4gkp}

\subsubsection{From physical LLRs to ${\rm C_4}$ logical LLRs}
\label{sec:physical_to_c4_llr}
Let
\begin{equation}
\ell_i \equiv \log \frac{\Pr(e_i=0\mid y_i)}{\Pr(e_i=1\mid y_i)}
\end{equation}
denote the physical GKP-derived log posterior odds for a Pauli error variable $e_i\in\{0,1\}$. Here $y_i$ is the accessible local measurement record for mode $i$, and the four-index set $i=1,\ldots,4$ refers to physical modes in one ${\rm C_4}$ block, not four child resources. When physical-error priors are uniform, these odds coincide with the likelihood-ratio convention in Eq.~\eqref{eq:gkp_llr_main}; with unequal priors the prior-odds term must be retained. Conditioning on the measured ${\rm C_4}$ syndrome $\bm s$, the logical log odds are obtained by marginalizing over physical error patterns consistent with each logical class,
\begin{equation}
\ell_{\Cfour}
=\log
\frac{\sum\limits_{\bm e\in\mathcal E_0(\bm s)}
\exp\!\left[\frac12\sum_{i=1}^{4}(1-2e_i)\ell_i\right]}
{\sum\limits_{\bm e\in\mathcal E_1(\bm s)}
\exp\!\left[\frac12\sum_{i=1}^{4}(1-2e_i)\ell_i\right]},
\label{eq:c4_llr}
\end{equation}
where $\mathcal E_{0/1}(\bm s)$ denotes the syndrome-compatible physical error patterns in each binary data-parity hypothesis for the selected Pauli component. No hard physical-level decision is required: every compatible pattern contributes with its likelihood weight. The resulting $\ell_{\Cfour}$ is passed to the next concatenation level.

\subsubsection{Parity-check structure}

The ${\rm C_4}$ stabilizers impose parity constraints on the physical error variables, so Eq.~\eqref{eq:c4_llr} can be viewed as inference on a small factor graph with physical LLRs as incoming messages. The same structure repeats at higher concatenation levels.

Higher-level stabilizers impose analogous constraints among ${\rm C_4}$ subblocks. Decoder-confidence postselection ranks the resulting reliability information and applies the fixed retained fraction $q_L$ while preserving the probabilistic interpretation across levels.

The same logical LLRs can be incorporated into adaptive MBQC measurement and frame updates as described in Sec.~\ref{app:softmbqc}.

\subsection{Logical-operation maps and gauge-qubit tracking}
\label{app:ds_mapping}
The physical permutation $\operatorname{SWAP}_{23}$ exchanges the two ${\rm C_4}$ logical-subsystem ports,
\begin{equation}
 (a_D,a_G)_{\rm in}\longmapsto(a_G,a_D)_{\rm physical\ ports}.
\label{eq:ds_swap_port_map}
\end{equation}
Consequently, a data label that follows the quantum information through the declared address map satisfies $D_{\rm in}\mapsto G_{\rm out}$, whereas a fixed physical $D$ port after the permutation contains the input gauge degree of freedom. The designated-$D$ certificate therefore transfers to the address-tracked output data label, not to an unrelabeled fixed port. The gauge qubit remains in the joint decoder state but is not independently qualified as a delivered output.
No statistical independence between the ${\rm C_4}$ data and gauge degrees of freedom is assumed.

For two ${\rm C_4}$ blocks $A$ and $B$, corresponding physical CNOTs implement
\begin{equation}
 \operatorname{CNOT}^{\otimes4}_{A\rightarrow B}
 =\overline{\operatorname{CNOT}}_{D_A\rightarrow D_B}
  \otimes\overline{\operatorname{CNOT}}_{G_A\rightarrow G_B}.
\label{eq:ds_cnot_map}
\end{equation}
This ideal map keeps the designated and gauge qubits separate and fixes their address-matched pairing.

\subsection{Logical output and residual-Pauli normalization}\label{app:c4_two_logical_subsystems}
Underlying each $[[4,1,1,2]]$ data--gauge realization is the $[[4,2,2]]$ stabilizer space. Designating the protected output as data $D$ and tracking $G$ as gauge determines the output-normalization convention; it does not change the four-qubit stabilizers. The recursive simulator retains the information needed to propagate both subsystems through the verification gadget. One pre-specified logical Bell-pair subsystem of the verified block pair is counted as the accepted encoded ancilla, while the gauge qubit is retained for parity consistency and optional diagnostics and is not credited as a second independent output.

After final deterministic frame composition, $a_L\in\{0,1,2,3\}$ labels the true Pauli class on the designated Bell pair and $\hat a_L$ is the decoder correction. The residual $e_L=\hat a_L\oplus a_L$ has accepted distribution
\begin{equation}
P_{\rm res,L}^{\rm prep}(e\mid A_L=1)
=\Pr[e_L=e\mid A_L=1],
\label{eq:primary_residual_channel}
\end{equation}

\begin{align}
P_{L}^{\rm PS,prep}(s_{\rm GKP};\mathbf q)
&=1-P_{\rm res,L}^{\rm prep}(0\mid A_L=1)
\nonumber\\
&=\Pr[\hat a_L\oplus a_L\neq0\mid A_L=1].
\label{eq:primary_canonical_failure}
\end{align}

The four-class residual distribution, not the maximum of component-wise Bell-bit errors, defines the primary metric. Joint tracking of data and gauge uses 16 Pauli classes for diagnostics only; the gauge is not credited as a second qualified output. Under the stated Pauli-mixture description, write $P_L^{\rm res}(e)=\Pr(e_L=e\mid A_L=1)$. The corresponding estimated Pauli-diagonal logical channel model (for ideal teleportation) is
\begin{equation*}
\Lambda_{\rm LR}^{(L)}(\rho)
=\sum_{e\in\{I,X,Y,Z\}}P_L^{\rm res}(e)\,
\overline P_e\rho\overline P_e^\dagger.
\end{equation*}
It is CPTP because the weights are nonnegative and sum to one. This representation is associated with the accepted Bell-resource distribution and ideal logical teleportation; it does not reconstruct coherent device noise or certify noisy online gates.

\section{Preparation, MN, and QRL noise models}\label{app:qrl_insertion_details}
This Appendix distinguishes the isotropic displacement channel used in the reported physical-coordinate simulations from gate-specific QRL variance formulas and an auxiliary factorized logical-Pauli approximation. No device-specific logical-channel fitting is used to obtain the headline error rates.

\subsection{Declared physical-coordinate evaluation}\label{sec:mbqc_insertion_noise}
The independent $15.6$-dB confirmation adds the specified displacements to physical GKP coordinates in the MN-on condition and performs final GKP correction in both conditions before logical decoding. This physical-coordinate evaluation differs from the auxiliary independent-channel approximation, which factorizes logical preparation and post-injection errors before XOR convolution. The latter is not used to replace the direct confirmation.

Let $\Omega_L^{\rm prep}$ be the joint state of the prepared quantum candidate and its classical record $Y$. The notation $\mathcal N_{\rm prep}^{(L)}$ elsewhere summarizes the GKP and finite-squeezing QND/SUM preparation-noise model; conditional preparation and selection are not assumed to be a single logical noise channel. Before final correction and decoding, the additional physical noise acts as
\begin{equation}
\Omega_L^{\rm post}
=\bigl(\mathcal N_{\rm QRL/MN}^{(L)}\otimes\mathrm{id}_{Y}\bigr)
 [\Omega_L^{\rm prep}].
\label{eq:noise_scope_summary}
\end{equation}
Here $\mathrm{id}_{Y}$ leaves the existing record unchanged. The added-noise operation is a mixture of physical displacement unitaries and is therefore CPTP; it is not the inferred logical Pauli distribution. Loss, detector inefficiency, mode mismatch, storage noise, and inter-operation correlations require additional calibrated terms. Controller latency, buffering, and throughput are not logical-error probabilities.

For the numerical results reported below, the two post-injection labels refer to one fixed pair of channels. In the \emph{MN-noise-off} condition no additional post-injection displacement is applied. In the \emph{MN-noise-on} condition a Gaussian displacement is drawn independently in each relevant quadrature. Its reference variance is evaluated at the numerical noise-quality level $s_{\rm MN}=s_{\rm GKP}+2$, i.e., $2\,\mathrm{dB}$ higher than $s_{\rm GKP}$, and multiplied by four. Equivalently, the declared added variance is
\begin{align}
\sigma_{q,{\rm MN,on}}^{2}=\sigma_{p,{\rm MN,on}}^{2}
&=4\,10^{-2/10}\sigma_{\rm GKP}^{2}(s_{\rm GKP}) \nonumber\\
&\simeq2.524\,\sigma_{\rm GKP}^{2}(s_{\rm GKP}).
\label{eq:frozen_mn_on_variance}
\end{align}
The isotropic numerical step is not identified with the gate-specific QRL parameterization below without a specified squeezing/gain and measurement-pattern calibration.

The second evaluation applies post-injection QRL/macronode noise after the logical Bell resource has been prepared. For each constituent qubit $a$ of the prepared level-$L$ resource we sample independent displacement noise
\begin{align}
 \hat q_a &\mapsto \hat q_a+\xi_{q,a},
&
 \hat p_a &\mapsto \hat p_a+\xi_{p,a},
\nonumber\\
 \xi_{q,a}&\sim\mathcal N(0,\sigma_{q,{\rm MN}}^2),
&
 \xi_{p,a}&\sim\mathcal N(0,\sigma_{p,{\rm MN}}^2),
\label{eq:independent_mn_noise}
\end{align}
where the variances are determined by the declared post-injection channel. For the headline MN-noise-on results they are the declared single-step values of Eq.~\eqref{eq:frozen_mn_on_variance}, while the MN-noise-off condition sets both additional variances to zero.

\subsection{Gate-specific QRL parameterization}

We use the QRL finite-squeezing parameterization of Ref.~\cite{MenicucciQRL2020}. Let $r_{\rm cl}$ be the squeezing parameter of the input squeezed resources used to generate the QRL cluster. Using the convention $[\hat q,\hat p]=i$ and $\mathrm{Var}(q_{\rm squeezed})=\tfrac12 e^{-2r_{\rm cl}}$, we define the corresponding input squeezing $s_{\rm cl}=(20/\ln 10)r_{\rm cl}$ in decibels. This cluster-source squeezing is not the effective GKP displacement-noise level $s_{\rm GKP}$ used in the headline data. The effective QRL edge weight and residual finite-squeezing parameter are
\begin{align}
 t_{\rm QRL} &= \tanh(2r_{\rm cl}),
&
 \epsilon_{\rm QRL} &= \operatorname{sech}(2r_{\rm cl}).
\label{eq:qrl_t_epsilon}
\end{align}
For a single QRL computation step, the quadrature noise factors are
\begin{align}
 N_x &= \tanh^{-2}(2r_{\rm cl}),
&
 N_p &= 1 .
\label{eq:qrl_noise_factors}
\end{align}
Reference~\cite{MenicucciQRL2020} uses $x,p$ quadratures; here we identify $x\equiv q$.

In the following gate labels, $I$ is the identity, $F$ is the single-mode Fourier transform, $P(1)$ is a unit-strength quadratic phase gate, and $CZ(1)$ is a unit-strength controlled-phase gate; the $q,p$ variance factors follow the chosen QRL convention. For the elementary QRL Clifford operations used here, the additive displacement variances are
\begin{align}
(v_q^I,v_p^I)&=\frac{\epsilon_{\rm QRL}}{2}(N_x,N_p),\\
(v_q^F,v_p^F)&=\frac{\epsilon_{\rm QRL}}{2}(N_x+N_p,N_x+N_p),\\
(v_q^{P(1)},v_p^{P(1)})&=\epsilon_{\rm QRL}(N_x,N_p),\\
(v_q^{(F\otimes F)CZ},v_p^{(F\otimes F)CZ})&=\frac{\epsilon_{\rm QRL}}{2}(N_x+N_p,N_x+N_p).
\label{eq:qrl_operation_variances}
\end{align}
These gate-dependent variance formulas follow Ref.~\cite{MenicucciQRL2020}; they are contextual analytic inputs, not a calibrated replacement for the isotropic noise used in the headline simulation.

\begin{table}[tbp]
\centering
\small
\begin{tabular}{lcc}
\toprule
QRL operation $g$ & $v_q^{g}$ & $v_p^{g}$ \\
\midrule
$I$ & $N_x\epsilon_{\rm QRL}/2$ & $N_p\epsilon_{\rm QRL}/2$ \\
$F$ & $(N_x+N_p)\epsilon_{\rm QRL}/2$ & $(N_x+N_p)\epsilon_{\rm QRL}/2$ \\
$P(1)$ & $N_x\epsilon_{\rm QRL}$ & $N_p\epsilon_{\rm QRL}$ \\
$(F\otimes F)CZ(1)$ & $(N_x+N_p)\epsilon_{\rm QRL}/2$ & $(N_x+N_p)\epsilon_{\rm QRL}/2$ \\
\bottomrule
\end{tabular}
\caption{Gate-dependent additive displacement variances in the QRL parameterization of this Appendix. These entries do not by themselves identify the device-level noise with the isotropic numerical step of Eq.~\eqref{eq:frozen_mn_on_variance}.}
\label{tab:supp_qrl_operation_noise}
\end{table}

\subsection{Auxiliary logical-Pauli factorization}
The headline MN-on study adds the prescribed displacement noise to physical GKP coordinates and then performs final correction and frozen decoding. For a separate \emph{auxiliary} approximation, let $e\in\{0,1,2,3\}$ be the preparation residual Pauli class and $u\in\{0,1,2,3\}$ an effective post-injection Pauli class on the designated Bell output. Their respective marginal distributions are denoted $P_{\rm res,L}^{\rm prep}(e\mid A_L=1,\mathbf q)$ and $P_{\rm QRL,L}(u\mid\mathcal S,h_{\rm route})$, where $\mathbf q$ is the declared retention schedule, $\mathcal S$ denotes an ordered QRL measurement/gate schedule, and $h_{\rm route}$ a routing history. The latter distribution is an auxiliary model input, not the output of a device-calibration fit performed in this work.
Under the factorized approximation, the residual Pauli class is the two-bit XOR convolution
\begin{align}
P_{\rm tot,L}(c)
=\sum_{e=0}^{3}&
P_{\rm res,L}^{\rm prep}(e\mid A_L=1,\mathbf q)
\nonumber\\[-2pt]
&\times P_{\rm QRL,L}(c\oplus e\mid\mathcal S,h_{\rm route}).
\label{eq:top_logical_use_convolution}
\end{align}
The corresponding auxiliary failure probability is
\begin{equation}
P_L^{\rm ind,post}=1-P_{\rm tot,L}(0).
\label{eq:independent_postinjection_error}
\end{equation}
Here $\oplus$ adds two-bit logical Pauli labels. This factorization assumes statistical independence between preparation and post-injection residual errors after conditioning on the accepted-output protocol. Independently drawn physical displacements alone do not ensure this logical independence under joint decoding; a correlated model requires their joint distribution. Accordingly, $P_L^{\rm ind,post}$ is not identified with the directly simulated physical-coordinate result $P_L^{\rm MN,post}$. The identity-pattern variances in Table~\ref{tab:supp_qrl_operation_noise} are generally anisotropic, unlike the prescribed isotropic MN-on variance in Eq.~\eqref{eq:frozen_mn_on_variance}; no calibrated equivalence between them is asserted.

\subsection{Ordered covariance propagation for a QRL segment}
Gate counts alone do not generally determine the output noise of a Gaussian circuit. Let $S_j$ be the linear quadrature map of step $j$ and $N_j$ the covariance of fresh additive noise at its output, both in the same specified quadrature convention. For independent noise increments in an ordered Gaussian segment~\cite{Weedbrook2012CVReview},
\begin{align}
\Sigma_j &= S_j\Sigma_{j-1}S_j^T+N_j,\\
\Sigma_{m}^{\rm added}&=\sum_{j=1}^{m}T_jN_jT_j^T,
\quad T_j=S_mS_{m-1}\cdots S_{j+1},\\
T_m&=I.
\label{eq:ordered_qrl_covariance}
\end{align}
The total covariance also includes $T_0\Sigma_0T_0^T$, with $T_0=S_m\cdots S_1$. The variance table gives diagonal entries only; a multimode segment also requires all relevant off-diagonal elements of $N_j$. This identity follows by substituting the preceding step into the next; it is not a new simulation result. A scalar sum of gate variances is a special case only if each contribution has already been transported into the same output frame, the relevant cross-covariances vanish, and the required diagonal entries are then selected. A shear or SUM can otherwise propagate and correlate earlier noise. Temporally correlated increments require the corresponding cross terms. GKP digitization, syndrome conditioning, and error-correction resets require their own conditional update rather than this unconditional Gaussian identity. Storage and routing noise must enter at their actual positions in the ordered sequence.

The covariance recursion is an auxiliary Gaussian propagation identity, not an evaluation performed for the reported MN-on condition. A compact gate-count record alone cannot replace the ordered maps, covariance blocks, and decoding/reset locations; no longer-circuit logical performance is inferred from that identity.

\subsection{Independent physical-noise model}

The independent-channel approximation is used to scan MN/QRL post-injection noise at low computational cost. 
This analysis assumes that the preparation-decoder logical Pauli error and the logical Pauli error induced by post-injection noise are independent. For the primary output subsystem, the MN/QRL post-injection noise is represented in this auxiliary diagnostic by a four-class logical Pauli-error distribution on the designated Bell-pair subsystem. When both ${\rm C_4}$ data/gauge degrees of freedom are retained jointly for diagnostics, the same construction can be lifted to a 16-class residual-error distribution corresponding to four binary frame components. The four-class convolution is Eq.~\eqref{eq:top_logical_use_convolution}, under the independence conditions of Appendix~\ref{app:qrl_insertion_details}. In this auxiliary model, $P_L^{\rm ind,post}(\bm q)=1-P_{\mathrm{tot},L}(0)$; it is not a substitute for the physical-coordinate result.

For the independent physical error model, each physical GKP coordinate receives MN noise with variance
\begin{equation}
 \sigma_{\MN,{\rm aux}}^2=C_{\MN}\sigma_{\GKP}^2.
\end{equation}
The physical bit-flip probability is approximated by
\begin{equation}
 p_{\mathrm{phys}}
 =
 \Pr\left(|X|>\frac{\sqrt{\pi}}{2}\right),
 \qquad
 X\sim{\cal N}(0,C_{\MN}\sigma_{\GKP}^2).
 \label{eq:app_physical_flip}
\end{equation}
For a level-\(L\) ${\rm C_4}$ block containing \(4^L\) physical qubits in the simplified counting model, the parity-flip probability is
\begin{equation}
 p_{\mathrm{par},L}
 =
 \frac{1-(1-2p_{\mathrm{phys}})^{4^L}}{2}.
 \label{eq:app_parity_flip}
\end{equation}
The four binary frame components ($X/Z$ on the data and gauge degrees of freedom of the ${\rm C_4}$ block) are then combined into the joint 16-class diagnostic distribution $P_{\MN,L}(\act)$ by treating the four parity flips as independent within this factorized approximation. The declared MN-noise-on effective channel retains
\begin{equation}
C_{\MN}^{\rm prod}=4\times10^{-2/10}\simeq2.524,
\end{equation}
consistent with Eq.~\eqref{eq:frozen_mn_on_variance}; MN noise off sets the additional variance to zero. The earlier sensitivity-family coefficients are not used in the reported headline results.

\section{Decoder records, selection-policy calibration, and measurement instruments}\label{app:decoder_details}
\subsection{Quantum-instrument composition at the online boundary}\label{app:instrument_composition}
An accepted encoded Bell pair is an ancilla distinct from the teleported data. Let $\mathsf D_L$ denote the encoded data and its classical address and correction records, $\mathsf B_L$ the accepted Bell resource and associated records, and $\mathsf C_L$ the newly produced measurement records. These registers are distinct from the single-candidate classical decoder record $\mathcal R_L$ and the nominal supplied-GKP resource cost $R_L^{\rm GKP}$. Injection $\mathcal J_L$ and the QRL measurement instrument $\mathcal M_{\rm QRL}$ are followed by classical processing and prescribed conditional corrections $\mathcal D_L^{\rm cq}$. Their composition is
\begin{equation}
\begin{aligned}
\mathcal I_L:&\quad \mathsf D_L\otimes \mathsf B_L
\longrightarrow \mathsf D_L\otimes\mathsf C_L,\\
&\quad\mathcal I_L=\mathcal D_L^{\rm cq}\circ\mathcal M_{\rm QRL}\circ\mathcal J_L.
\end{aligned}
\label{eq:interface_cycle}
\end{equation}
The complete quantum/classical output defines a quantum instrument. The classical-processing map implements the specified outcome-conditioned corrections without assuming that an inferred Pauli-error class is correct. This expression describes composability at the interface and is not evaluated as a complete noisy logical-gate channel in the present numerical study.

The main text defines classical decoder updates and binary candidate selection. For the quantum instrument in Eq.~\eqref{eq:interface_cycle}, an outcome $y$ labels a linear completely positive, trace-nonincreasing operation $\mathcal E_y$; summing (or integrating) over all outcomes is trace preserving. Recording the outcome gives $\sum_y\mathcal E_y(\rho)\otimes|y\rangle\langle y|$. Normalizing one accepted branch is a conditional state update, not generally a linear quantum channel. Fixed-batch rank selection is a classical rule on the joint batch record, not a fixed independent single-candidate instrument. This Appendix specifies the logical endpoint, fitted-map inputs, deterministic transformations, and auxiliary diagnostics.

\paragraph*{Class-score construction and calibration.}
The decoder uses two side-specific ExtraTrees predictors to produce a normalized 64-class score distribution, $p_A(u\mid y)$ and $p_B(v\mid y)$. Absent classes receive a small positive numerical floor before normalization. The four Bell-Pauli scores are constructed by summing products of the side scores over pairs of component labels consistent with the fixed Bell-stabilizer quotient map $\lambda(u,v)\in\{0,1,2,3\}$ and composing the recorded Pauli frame $f_L^{\rm known}$:
\begin{equation}
\begin{aligned}
 s_L(a\mid y)&=\sum_{u,v:\,\lambda(u,v)=a}p_A(u\mid y)p_B(v\mid y),\\
 p_L(a\mid y)&=s_L(a\oplus f_L^{\rm known}\mid y).
\end{aligned}
\label{eq:frozen_decoder_class_scores}
\end{equation}
The product combines two fitted side-score distributions and need not equal a true jointly calibrated conditional posterior. No separate post-hoc probability-remapping function enters this score construction: class normalization and known-frame relabeling are deterministic operations, not temperature, isotonic, or sigmoid calibration. Instead, non-test calibration data are used to select the risk/retention policy. The estimated score is not claimed to be a reliability-validated conditional failure probability. The isotropic MN benchmark does not infer a device-specific logical error distribution.
\subsection{Pre-specified recursive-decoder definition}
\label{app:frozen_recursive_decoder_contract}
The frozen level-2 models are \texttt{ExtraTreesClassifier} ensembles with 112 trees, maximum depth 28, minimum leaf size 3, square-root feature subsampling, and balanced-subset class weights. Each level-2 MN-specific model bundle accepts 523 ordered observable/derived input features and contains the two side-specific multiclass estimators used in Eq.~\eqref{eq:frozen_decoder_class_scores}. Model training uses data disjoint from the independent confirmation samples; at the reported $q=1$ level-2 point, no decoder-confidence cutoff removes candidates. The feature definition and fitted parameters are fixed before confirmation, with no retraining or retrospective model selection. For each child $k\in\{1,\ldots,12\}$ supplied by the preparation gadget, the parent decoder receives (i) normalized logical-class-score information after deterministic Pauli-frame handling, (ii) exposed ${\rm C_4}$ verification syndromes and Bell-measurement outcomes, (iii) ordered child-slot/address information, and (iv) retained analog GKP likelihood features. No unobservable physical error label is supplied to the decoder during evaluation.

\paragraph*{Frozen evaluation inputs.}
The decoder's operational input contains measured GKP analog likelihoods, exposed verification syndromes and Bell outcomes, the known frame, and child-slot/address records. Simulated correct labels are used for fitting and non-test policy comparison only. Neither those labels nor hidden physical-error coordinates are supplied as decoder features at independent evaluation. The fixed feature columns, learned trees, logical-class map, and risk-ranking rule are applied without updating the model from confirmation outcomes.

\subsection{Selection-policy calibration and independent evaluation}\label{app:decoder_policy_calibration}
The evaluation separates three data roles. The training set $\mathcal T_{\rm fit}$ fits the component multiclass estimators using observable records and simulated labels. Non-test calibration data $\mathcal T_{\rm sel}$ compare the specified decoder/risk and selection-policy candidates using selected-output failure statistics and retain the admissible policy, including the level-dependent schedule $\mathbf q^\star$. The independent set $\mathcal T_{\rm conf}$ evaluates the fixed choice; its labels are not used for refitting, policy reselection, or choosing an operating point after the fact. At $q_2=1$ every candidate is retained, so the level-2 confirmation tests the frozen class decision rather than the benefit of confidence filtering.

For each generated candidate $i$ at level $L$, the fixed decoder returns $p_{L,i}(a\mid y_i)$ and the associated score
\begin{equation}
\begin{aligned}
\hat a_{L,i}&=\arg\max_{a\in\{I,X,Y,Z\}}p_{L,i}(a\mid y_i),\\[-2pt]
r_{L,i}&=1-\max_a p_{L,i}(a\mid y_i).
\end{aligned}
\label{eq:policy_calibration_risk}
\end{equation}
The selection policy reported for the $12.25$-dB level-2--4 ordering study is \emph{within-batch rank selection}. For a batch $B$ containing $M$ generated candidates, let $\rho_B(i)$ be the rank of $(r_{L,i},\mathrm{jobID}_i,\mathrm{shotID}_i)$ in ascending lexicographic order; the identifiers break score ties deterministically. The acceptance indicator is
\begin{equation}
\begin{aligned}
K_L(B)&=\max\{1,\lceil q_L^\star M\rceil\},\\[-2pt]
A_{L,i}(B)&=\mathbf 1[\rho_B(i)\le K_L(B)].
\end{aligned}
\label{eq:policy_calibration_rank}
\end{equation}
Here $M=500$ for the stated level-2--4 batches, and $(q_2,q_3,q_4,q_5)=(1,0.70,0.96,1)$ is the preselected schedule. This policy imposes an exact rank quota within each specified batch, not an absolute threshold on $r_L$ or a demonstrated streaming implementation. The level-5 condition retains all candidates; the separate $15.6$-dB level-2 confirmation also has $q_2=1$ and therefore no confidence-based rejection.

One diagnostic that can be constructed using only calibration labels is the empirical selected-output error rate for a proposed policy $q$,
\begin{equation}
\widehat P_{L,\rm sel}^{\rm cal}(q)
=\frac{\sum_{i\in\mathcal T_{\rm sel}}A_{L,i}(q)\,\mathbf 1[\hat a_{L,i}\ne a_{L,i}]}
{\sum_{i\in\mathcal T_{\rm sel}}A_{L,i}(q)},
\label{eq:policy_calibration_diagnostic}
\end{equation}
whenever the denominator is nonzero. This is a non-test selected-output statistic used for policy comparison; it is not a probability-remapping objective and is never reported as the independent logical-error estimate. Candidate operating points and nominal costs are selected according to the declared non-test rule of Appendix~\ref{app:resource_accounting}. After the policy is frozen, the same formula with $\mathcal T_{\rm conf}$ in place of $\mathcal T_{\rm sel}$ defines the \emph{observed selected-output failure fraction}; the confidence/credible intervals are evaluated from confirmation counts using their separately stated statistical conventions (Appendix~\ref{app:statistics}). The true class $a_{L,i}$ is used only in labeled training, non-test calibration diagnostics, and independent evaluation---never as a runtime decoder feature or selection key.

An optional streaming fixed-cutoff policy would set an acceptance threshold from non-test score quantiles; it was not used for the reported $12.25$-dB batchwise selection. No calibration-only estimate, however accurate, is substituted for the independent reported error counts. In particular, an empirical agreement of predicted class confidences with class frequencies is a separate reliability test and is \emph{not} established by the rank-selection equations above.

The twelve children are placed in the fixed ordered groups defined by the Knill-type parent verification gadget. Deterministic ${\rm C_4}$ parity constraints, Bell-stabilizer quotienting, child-slot permutations, and Pauli-frame composition are evaluated exactly. The statistical model is used only to estimate normalized scores over the four logical Pauli classes after these exact transformations. The fitted map cannot alter the code relations, logical labels, address map, or deterministic frame composition. Decoder-universal robustness is not inferred from the fixed-map result. In particular, the decoder output is
\begin{equation}
\begin{aligned}
p_{L+1}(a\mid y)
=\mathcal F_{L+1}\!\bigl[
&\{p_{L,k},s_{L,k},u_{L,k},A_{{\rm mode},L,k}\}_{k=1}^{12},\\
&s_{L+1},u_{L+1}\bigr],
\end{aligned}
\label{eq:supp_frozen_recursive_map}
\end{equation}
where $s$ denotes the observable syndrome/verification record, $u$ collects the pre-specified analog-likelihood features, and $A_{\rm mode}$ denotes the ordered address/slot record. The tracked Pauli frame after the final GKP correction is then composed exactly with this class-score distribution. The predicted class is $\hat a_{L+1}=\arg\max_a p_{L+1}(a\mid y)$ and the candidate-ranking score is
\begin{equation}
r_{L+1}=1-\max_a p_{L+1}(a\mid y).
\label{eq:supp_decoder_uncertainty}
\end{equation}
The $12.25$-dB evaluation ranks this score within each replicate as in Appendix~\ref{app:selection_policy}. If the final GKP correction supplies a deterministic Pauli-frame update $f_{\rm GKP}^{\rm final}$ (the known update from this final correction), the post-correction class-score distribution is obtained exactly as
\begin{equation}
p_{\rm post}(a\mid y,f_{\rm GKP}^{\rm final})
=p_{\rm pre}(a\oplus f_{\rm GKP}^{\rm final}\mid y),
\label{eq:known_frame_composition}
\end{equation}
so the applied correction is never presented to the statistical model as an unknown fault.

The equations specify the observable input classes and deterministic transformations used by the decoder. The reported performance pertains to the fitted decoder fixed for each evaluation; it does not imply robustness to arbitrary decoder replacements. Level-dependent fitted maps are used where required by the recursive construction.

The parent-level class scores $p_{L+1}(a\mid y_{L+1})$ are the class-score component of the recursive record update in Eq.~\eqref{eq:recursive_soft_decoder_map}, with inputs from the twelve child records and newly accessible likelihood, syndrome, address, and frame data.

\subsection{Physical-coordinate-level decoder}

The physical-coordinate-level MN model adds noise to the GKP coordinates before logical decoding. It is distinct from the independent-channel approximation and is used for the independent $15.6$-dB confirmation. For each resource, the coordinates before the additional noise step are retained, and MN/QRL noise is sampled at the physical-coordinate level,
\begin{equation}
 x_j^{\mathrm{noisy}}
 =
 x_j+\xi_j,
 \qquad
 \xi_j\sim{\cal N}(0,\sigma_{\MN}^2),
\end{equation}
and similarly for the conjugate quadrature when required by the measured channel. The noisy physical bit is obtained by nearest-lattice binning,
\begin{equation}
 b_j^{\mathrm{noisy}}
 =
 \left[
 \mathrm{round}\!\left(\frac{x_j^{\mathrm{noisy}}}{\sqrt{\pi}}\right)
 \right]\bmod 2 .
\end{equation}
In the physical-coordinate description, simulator coordinates define the noise experiment. Only measurement-accessible outcomes, syndromes, known frames, and their pre-specified derived features may be passed to an operational MN-aware decoder. Here $\Phi_{\rm obs}$ denotes the prescribed construction of the observable feature record, $\mathsf M_{\rm allowed}$ specifies the permitted measurements, and $\mathsf F_{\rm known}$ is the tracked correction record; the map is separate from the ${\rm C_4}$ gauge group $\mathcal G$. Let $\bm y_{\rm obs}^{\rm noisy}=\Phi_{\rm obs}(\bm x^{\rm noisy};\mathsf M_{\rm allowed},\mathsf F_{\rm known})$ denote the decoder-input record formed from permitted measured outcomes, their declared derived features, and known classical frames. The symbol $\bm x^{\rm noisy}$ describes simulator coordinates and is not itself the operational decoder input:
\begin{equation}
\hat{\act}=D_{\MN}^{(L)}(\bm y_{\rm obs}^{\rm noisy}).
\end{equation}
The operational interpretation of this map requires the features constructed by $\Phi_{\rm obs}$ to depend only on measured quantities and known classical records, not on hidden simulator labels or unmeasured quadratures.

The physical-coordinate-level decoder is not required to be the same as the preparation decoder. In particular, the preparation decoder predicts the residual logical Pauli error of an offline resource, whereas the MN-aware decoder predicts the residual logical Pauli error after post-injection noise. These two tasks have different feature distributions.

The designated-data identity Pauli class dominates in the low-error regime. Aggregate selected-output error rates alone do not separately quantify recall for each rare nonidentity Pauli class; no such per-class sensitivity metric is claimed in the reported results.

\subsection{Diagnostic comparison with a no-residual-inference baseline}
\label{app:decoder_ablation}
The frozen aggregate records permit one diagnostic comparison without retraining or reselecting a decoder. An identity residual predictor applies the deterministic Pauli-frame update but performs no statistical inference and therefore fails whenever the exact designated-$D$ residual Pauli operation is nontrivial. Table~\ref{tab:decoder_ablation} compares this no-residual-inference baseline with the fixed four-class decoder on the same $q=1$ independent samples.

\begin{table*}[t]
\centering
\footnotesize
\setlength{\tabcolsep}{6pt}
\begin{tabular*}{\textwidth}{@{\extracolsep{\fill}}p{0.26\textwidth}p{0.22\textwidth}p{0.22\textwidth}p{0.16\textwidth}}
\toprule
Inference rule & No additional macronode-noise step & One macronode-noise step & Interpretation \\
\midrule
Identity residual predictor after deterministic frame update & $9/10^6$; $U_{0.95}=1.5705\!\times\!10^{-5}$ & $6163/10^6$; $U_{0.95}=6.2933\!\times\!10^{-3}$ & No statistical residual-Pauli inference \\
Fixed constrained four-class decoder & $3/10^6$; $U_{0.95}=7.7536\!\times\!10^{-6}$ & $4/10^6$; $U_{0.95}=9.1535\!\times\!10^{-6}$ & Primary independent result \\
\bottomrule
\end{tabular*}
\caption{Diagnostic comparison at $15.6$ dB and $q=1$. Both rows use the same frozen independent samples. The identity predictor is a no-residual-inference baseline. This diagnostic isolates the contribution of residual-Pauli inference and is not used to define the confirmatory confidence bound.}
\label{tab:decoder_ablation}
\end{table*}

\section{Numerical evaluation, statistical confidence, and resource accounting}\label{app:numerical_resource_protocol}
\subsection{Selection and confirmatory statistics}\label{app:statistics}
\subsubsection{Fixed protocol, selection policy, and output definition}\label{app:selection_policy}
The confirmatory data are generated only after the decoder family, observable feature set, logical-Pauli definition, the primary level-dependent retention schedule $\mathbf q^{\star}$, and the confidence criterion have been fixed. Training, calibration, schedule selection, and independent evaluation use disjoint replicate or seed-bank sets. The primary schedule is selected on calibration data and is not reselected on the independent test set.

Candidates are generated at each specified level and displacement-noise setting and decoded after final GKP correction. In the soft-verification mode, syndrome outcomes are recorded without an additional hard-rejection test. At $12.25$ dB, each level-2--4 replicate contains $M=500$ generated candidates. They are sorted by risk, with ties ordered by job and shot identifiers; the first $\max\{1,\lceil Mq_L\rceil\}$ are retained. This batchwise rank rule requires scoring and comparing the specified candidate set; associated candidate buffering and selection latency are outside the nominal supplied-GKP count and protected online layer depth. A separate fixed-cutoff streaming rule is discussed only as an alternative and is not assumed in these $M=500$ results. True-error labels are not sorting keys. The $72$ replicates per MN condition give $36{,}000/36{,}000$, $25{,}200/36{,}000$, and $34{,}560/36{,}000$ selected/generated outputs at $L=2,3,4$. Level 5 retains $15{,}000/15{,}000$ at $q_5=1$. These are retention fractions at the modeled candidate boundary, not measured optical-source success probabilities.

Selection follows recursive message construction. The retained lower-level output sets are not fed back into generation of the higher-level test candidates. Thus these data do not characterize a factory with the same confidence filters cascaded through every internal child. This distinction concerns confidence rejection, not omission of verification measurements. The $15.6$-dB $q=1$ confirmation retains every generated candidate.

For each level and operating point, the attempted count, accepted count, logical-failure count, acceptance, logical Bell-pair error estimate, and uncertainty interval are retained. No decoder parameter, feature definition, logical-output definition, decoder-risk definition, or retention fraction is reselected from the confirmatory labels.

\begin{figure*}[t]
\centering
\includegraphics[width=1\textwidth]{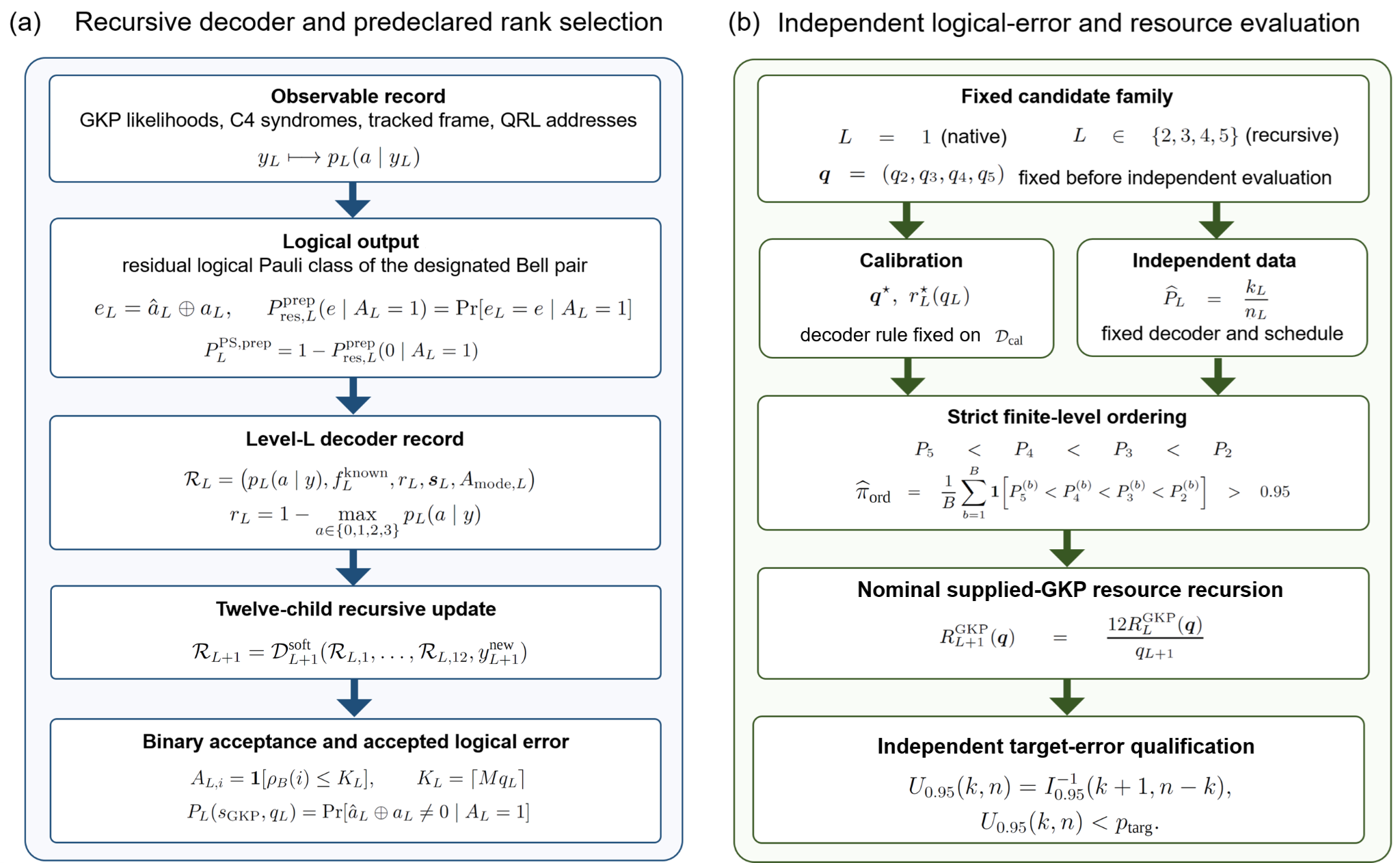}
\caption{Recursive decoding and independent evaluation. (a) Observable analog information, ${\rm C_4}$ syndromes, known frames, and mode addresses determine the designated Bell-pair class-score distribution and reliability record. Twelve child records and new verification data produce the parent record; acceptance remains binary [Eqs.~\eqref{eq:recursive_soft_resource_record}--\eqref{eq:soft_information_conditional_error}]. (b) Decoder parameters and the retention schedule are fixed before independent evaluation. The posterior ordering statistic $\widehat\pi_{\rm ord}$ is defined in Appendix~\ref{app:statistics}; nominal supplied-GKP cost and calibration-only candidate selection are defined in Appendix~\ref{app:resource_accounting}. The displayed Clopper--Pearson expression is the convention for the fixed-sample target confirmation. Panel (a) shows the predeclared batchwise rank-selection rule used in the numerical evaluation; true logical-error labels are used only to evaluate failures, not as decoder inputs. The boxes do not introduce fractional outputs or an observed retry-cost estimator.}
\label{Evaluation}
\end{figure*}

\subsubsection{Jeffreys-posterior ordering statistic and evaluated grid}
For each level $L$, let $k_L$ be the observed logical-failure count among $n_L$ accepted outputs. The aggregate failure-rate parameter $p_L$ used here is distinct from the candidate-level decoder class-score distribution $p_L(a\mid y)$ and its decision-risk score $r_L$. Under the declared binomial model with the Jeffreys prior, the level-wise posterior is
\begin{equation}
p_L\mid k_L,n_L\sim\operatorname{Beta}(k_L+1/2,n_L-k_L+1/2).
\end{equation}
The independent ordering evaluation draws these level-wise posteriors independently and estimates
\begin{equation}
\widehat\pi_{\rm ord}=\frac{1}{B}\sum_{b=1}^{B}
\mathbf 1\!\left[p_5^{(b)}<p_4^{(b)}<p_3^{(b)}<p_2^{(b)}\right].
\end{equation}
For the reported $12.25$-dB result, the evaluation uses $B=900{,}000$ posterior draws. The resulting estimates are approximately $0.99998$ (MN off) and $0.99981$ (MN on). These are Monte Carlo estimates conditional on the independent-Jeffreys posterior model, not exact probabilities; their finite-draw standard errors are of order $5\times10^{-6}$ and $1.5\times10^{-5}$, respectively, under independent posterior draws. The criterion used for the evaluated-grid diagnostic is $\widehat\pi_{\rm ord}>0.95$. This ordering statistic uses a declared independent-Jeffreys model for aggregate held-out failure counts; it is neither a paired cluster bootstrap nor an estimate of cross-level covariance from shared ancestry. The pointwise Jeffreys intervals in Fig.~\ref{fig:fresh_recursive_contraction} are separate summaries. The finite-level ordering claim is therefore conditioned on the declared binomial and level-wise posterior-independence model. As a descriptive split-lane check, each of the three independent lower-level lanes separately satisfies $P_4<P_3<P_2$ in both MN conditions, and each of the six level-5 lanes has zero failures in $2{,}500$ selected roots per condition. Within each lower-level lane, rank selection is performed separately for each replicate. It couples acceptance indicators within a batch; independent posterior draws do not establish independent selected errors or correct for shared resource ancestry. The diagnostic therefore neither supplies a covariance-corrected guarantee nor establishes equivalence to a fixed-cutoff stream.

The primary schedule $\mathbf q^\star$ is fixed on calibration data before the independent ordering sample is opened (Appendix~\ref{app:decoder_policy_calibration}). For the finite set $\mathcal G_{\rm grid}$ of evaluated GKP displacement-noise levels (not the gauge group $\mathcal G$ of Appendix~\ref{sec:concat_c4_stabilizers}), the first point meeting the criterion in both conditions is
\begin{align}
s_{\rm rec,grid}^{\star}(\mathbf q^{\star};\mathcal G_{\rm grid})
=\min\bigl\{s\in\mathcal G_{\rm grid}:\;&
\widehat\pi_{\rm ord}^{\rm off}(s)>0.95,\nonumber\\[-2pt]
&\widehat\pi_{\rm ord}^{\rm on}(s)>0.95\bigr\}.
\label{eq:finite_level_ordering_criterion}
\end{align}

\begin{figure*}[t]
\centering
\includegraphics[width=0.90\textwidth]{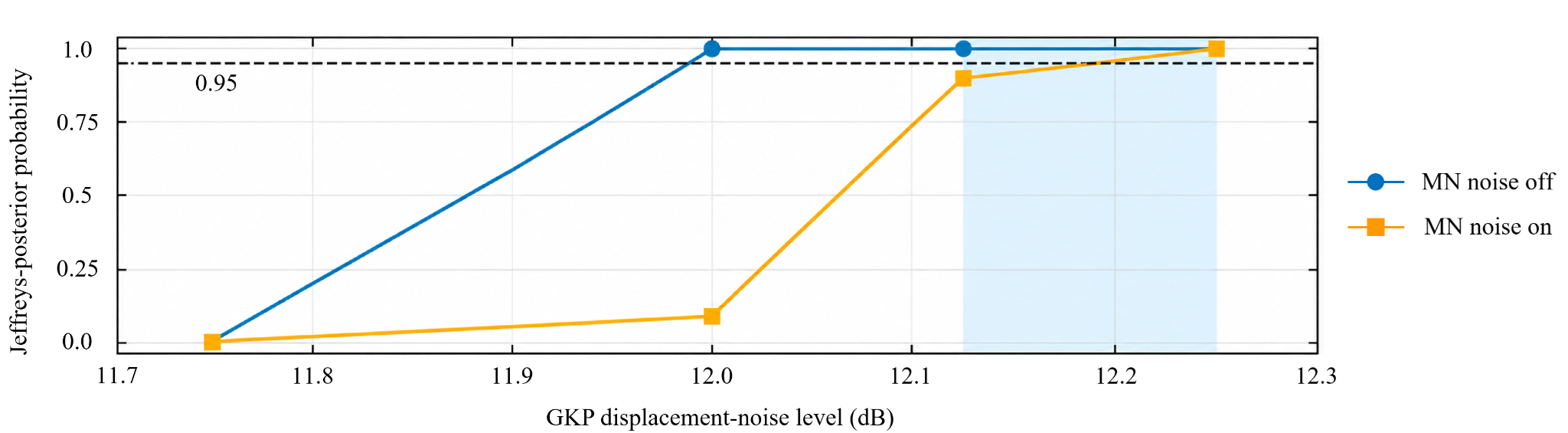}
\caption{Jeffreys-posterior probability of the strict ordering $P_5<P_4<P_3<P_2$ for the pre-specified postselection schedule. The dashed line is the $0.95$ criterion. The $12.25$-dB point is the first evaluated point at which both conditions exceed the $0.95$ criterion; the adjacent $12.125$-dB point does not. This evaluated-grid transition is used to support the finite-level ordering and is not a standard physical--logical pseudothreshold.}
\label{fig:ordering_probability_appendix}
\end{figure*}

The adjacent points give Eq.~\eqref{eq:recursive_grid_transition_interval}. Figure~\ref{fig:ordering_probability_appendix} shows this ordering diagnostic over the evaluated GKP displacement-noise grid for both specified MN conditions. A candidate family $\mathcal Q$ may be scanned on training/calibration data, but the independent test uses only the fixed schedule. No continuous threshold is inferred by interpolation.

\subsubsection{Pointwise intervals and fixed-sample qualification}
For the Jeffreys convention, a binomial likelihood and the prior $p\sim\operatorname{Beta}(1/2,1/2)$ give the posterior $p\mid k,n\sim\operatorname{Beta}(k+1/2,n-k+1/2)$. Writing $Q_u(a,b)$ for a beta-distribution quantile, the one-sided upper limit and central interval are
\begin{align}
U^{\rm J}_{0.95}(k,n)&=Q_{0.95}(k+1/2,n-k+1/2),\\
I^{\rm J}_{0.95}(k,n)&=[Q_{0.025}(k+1/2,n-k+1/2),\nonumber\\
&\hspace{1em}Q_{0.975}(k+1/2,n-k+1/2)].
\label{eq:jeffreys_conventions}
\end{align}
These are posterior credible statements, not Clopper--Pearson coverage guarantees. In particular a $95\%$ one-sided limit is not the upper endpoint of a central $95\%$ interval. Here $n$ is the accepted-sample count to which the binomial likelihood is applied, not an interchangeable raw-trial count.

For a binomial failure count $k$ in $n$ fixed selected samples, the one-sided Clopper--Pearson upper confidence limit is
\begin{equation}
U_{1-\alpha}(k,n)=I^{-1}_{1-\alpha}(k+1,n-k),
\label{eq:cp_upper}
\end{equation}
where $I^{-1}$ is the inverse regularized incomplete beta function, for $0\le k<n$; for $k=n$ the upper limit is $1$. The stated coverage is conditional on a fixed-$n$ binomial model: the trial-level failure indicators must have the relevant independent, common-probability distribution after fixing the decoder and acceptance rule. Distinct seeds or a large aggregate $n$ do not by themselves establish this condition. Shared resource ancestry, within-batch quota selection, and outcome-dependent stopping require separate treatment. The independent $15.6$-dB evaluation uses a fixed sample size with no result-dependent extension; diagnostic samples are not pooled into that evaluation. At this $q=1$ operating point, no confidence-based batch rank rejection is applied. The intended binomial unit is the accepted designated Bell-pair output, not an individual GKP measurement or child resource. The binomial coverage statement additionally assumes independent root-level failure indicators, including no sharing of child-resource ancestry across those trials; this condition cannot be inferred from the fixed sample count alone. For the independent $15.6$-dB confirmation, the two specified macronode-noise conditions are reported separately: at $L=2$ and $q=1$, the counts are $3/10^6$ and $4/10^6$, with upper limits $7.7536\times10^{-6}$ and $9.1535\times10^{-6}$. No familywise $95\%$ statement across both conditions is made.

\subsubsection{Separate level-5 curve-support samples}\label{app:curve_support}
At $11.75$ dB, the level-5 curve-support markers correspond to zero observed failures among $2{,}500$ accepted samples in each MN condition. These samples are separate from those used for the finite-level ordering test.
Figure~\ref{fig:recursive_contraction_results} includes the additional level-5 curve-support calculation at $12.00$ dB. With the extra post-injection macronode-noise step off, no failure was observed in $200{,}000$ accepted samples, giving a one-sided $95\%$ Jeffreys upper credible limit of $9.60\times10^{-6}$. Under the exact-binomial criterion used for the independent $15.6$-dB certificate, the corresponding one-sided $95\%$ Clopper--Pearson upper bound is $1.50\times10^{-5}$ and therefore does not qualify the $10^{-5}$ target. With the step on, the result is $5/200{,}000=2.5\times10^{-5}$, with a $95\%$ Jeffreys interval of $9.54\times10^{-6}$ to $5.48\times10^{-5}$. These samples support the plotted level-5 trend only. They neither replace the posterior ordering analysis used for the evaluated-grid transition interval nor constitute the independent exact-binomial $10^{-5}$ certificate of Sec.~\ref{sec:direct_target_qualification}. To keep the scan and confirmatory roles distinct, level-5 points at and above $12.125$ dB are not included in Fig.~\ref{fig:recursive_contraction_results}; the independent $12.25$-dB ordering sample remains in Fig.~\ref{fig:fresh_recursive_contraction}.

\subsubsection{Numerical records for resource analysis}\label{app:numerical_output_records}
The preparation and post-injection evaluations retain the following records for the resource analysis. Here ${\rm CI}_L^{\rm prep}$ denotes the reported preparation-error interval, $N_L^{\rm acc}$ the accepted-output count, and $N_L^{\rm fail}$ the designated-output logical-failure count. The preparation-only dataset is
\begin{align}
\mathcal D_{\rm prep}
=
\bigl\{&
L,\,
s_{\rm GKP},\,
\mathbf q,\,
P_{L}^{\rm PS,prep},
\nonumber\\
&
{\rm CI}_{L}^{\rm prep},\,
P_{\rm acc}^{(L)},\,
N_L^{\rm acc},\,N_L^{\rm fail}
\bigr\}_{L=2}^{5}.
\label{eq:dataset_prep_recursive}
\end{align}
For the reported post-injection evaluation, the records are indexed by $L$, $s_{\rm GKP}$, the specified MN-on/off condition, and the fixed retention schedule $\mathbf q$; the selected-output error counts and uncertainty intervals are reported for each condition. Cluster squeezing $s_{\rm cl}$, QRL schedule $\mathcal S$, routing history $h_{\rm route}$, and an auxiliary logical Pauli distribution $P_{\rm QRL,L}$ belong only to the separate gate-specific model of Appendix~\ref{app:qrl_insertion_details}; they are not independently calibrated inputs to the reported effective MN displacement step.
The physical-coordinate confirmation is distinct from the auxiliary independent logical-Pauli approximation described in Appendix~\ref{app:qrl_insertion_details}.

\subsection{Nominal and expected resource accounting}\label{app:resource_accounting}
\subsubsection{Declared retention cost}
The normalization is one accepted designated data-subsystem Bell resource, not two separately qualified ${\rm C_4}$ outputs. The supplied-GKP metric counts the preparation and rejection costs assigned by the recurrence below. It does not combine simultaneous optical modes, detectors, switches, delay lines, squeezed-vacuum ancillas, classical hardware, footprint, power, storage, and total execution time.

The finite-batch evaluation in Appendix~\ref{app:selection_policy} retains the prescribed rank quota $q_L$, and its accepted logical-error probability is the conditional probability in Eq.~\eqref{eq:soft_information_conditional_error}. Unlike a streaming threshold applied to unseen candidates, the declared protocol fixes the retained count within each scored batch. For a batch with fixed cost per generated candidate and $n$ retained outputs, the selection penalty is exactly $M/n$, here $1/q_L$. This output-stage ratio alone does not establish the cost of a recursively confidence-filtered child supply. Schedules and decoders are fixed before independent evaluation.

For the twelve-child factory, let $R_L^{\rm GKP}$ denote the supplied-GKP cost assigned by the declared schedule-level accounting convention. This nominal schedule metric counts supplied GKP qubits per designated Bell resource, not simultaneous hardware or a directly measured mean delivered cost. Actual mean consumption additionally requires the specified retry and supply process. With a declared baseline and nominal retained fraction $q_{L+1}$, its recursion is
\begin{equation}
R_{L+1}^{\rm GKP}(\mathbf q)
=\frac{12R_L^{\rm GKP}(\mathbf q)}{q_{L+1}}.
\label{eq:nominal_resource_recursion}
\end{equation}
Equation~\eqref{eq:nominal_resource_recursion} assigns the retention penalty $1/q_{L+1}$ at each level. It is not an observed internal-child retry ledger. No extra factor for a syndrome-only hard-rejection test is required by the reported soft-verification mode, since that test is not applied. This does not include experimental source-generation failures, storage loss, or finite-capacity waiting.

For the $12.25$-dB schedule, $\prod_{\ell=2}^{5}q_\ell=0.672$ and its inverse is $1.488095\ldots$. These summarize the nominal level-wise penalties; $67.2\%$ is not a measured end-to-end factory success probability. The measured output-retention fractions are given separately in Appendix~\ref{app:selection_policy}.

\subsubsection{Calibration selection and independent qualification}
For fixed GKP displacement-noise level $s_{\rm GKP}$, macronode-noise condition $m$, target $p_{\rm targ}$, and a pre-specified candidate family, selection and qualification are separate steps. Write $\theta=(L,\mathbf q)$ and suppress the fixed arguments of $R^{\rm nom}_{\rm GKP}$. Here $R^{\rm nom}_{\rm GKP}$ denotes the same declared schedule cost as $R_L^{\rm GKP}$ above. The rule is
\begin{align}
\theta_{\rm sel}
&\in\operatorname*{arg\,min}_{\theta\in\mathcal F_{\rm cal}}
R^{\rm nom}_{\rm GKP}(\theta),\nonumber\\
Q_{\rm conf}(\theta_{\rm sel})
&=\mathbf 1\!\left[
U_{0.95}^{\rm conf}(\theta_{\rm sel})\le p_{\rm targ}
\right].
\label{eq:target_resource_policy_selection}
\end{align}
Here $\mathcal F_{\rm cal}$ is the candidate subset admitted by the pre-specified calibration criterion, not a set filtered using independent confirmation outcomes. The selected candidate is fixed before the disjoint confirmation data are opened; independent data determine only its qualification. If $Q_{\rm conf}=1$, the reported cost is $R^{\rm nom}_{\rm GKP}(\theta_{\rm sel})$; otherwise that test does not qualify the selected candidate, and it does not trigger retrospective reselection. The bound $U_{0.95}^{\rm conf}$ uses the interval convention specified for that evaluation. The GKP displacement-noise level is held fixed because improving physical GKP quality is not counted as free by the supplied-GKP-qubit metric. The distinct diagnostic selection of the $15.6$-dB point and its new independent confirmation are described in Sec.~\ref{sec:direct_target_qualification}.

\subsubsection{Operating-point costs and source-quality break-even}
Using the same nominal schedule-cost convention, $R_{L=2}=288$ is the declared level-2 baseline and the level-3 point uses the pre-specified retained fraction $q_3=0.95$. The baseline is consistent with an algebraically implied level-1 nominal count $R_{L=1}=288/12=24$ and a twelve-child step with $q_2=1$. The value $24$ is an inferred normalization, not a verified component-by-component physical supply ledger; the breakdown of GKP states, qunaught inputs, and any non-counted ancillas must be checked against the factory circuit before identifying it with a complete source cost. Equation~\eqref{eq:nominal_resource_recursion} then gives
\begin{align}
R_{L=2}&=288,\\
R_{L=3}&=\frac{12\times 288}{0.95}=3637.8947,
\end{align}
and therefore
\begin{equation}
\frac{R_{L=3}}{R_{L=2}}=12.6316.
\label{eq:resource_cliff_ratio}
\end{equation}
The $14$-dB level-3 point meets the $10^{-5}$ target under the declared one-sided Jeffreys-posterior convention ($0/199{,}500$), whereas the $15.6$-dB level-2 point has a separate fixed-sample Clopper--Pearson qualification ($4/10^6$ for MN on). The $12.63$ nominal cost ratio therefore does not represent a common exact-binomial confidence certificate for both operating points. The supplied-qubit ratio does not assign equal generation cost to GKP qubits of different quality. Let $C_{\rm src}(s)$ denote the mean source cost per supplied GKP qubit at displacement-noise level $s$. Within this multiplicative source-cost model, the $15.6$-dB $L=2$ point has a lower assigned source cost than the $14$-dB $L=3$ point whenever
\begin{equation}
\frac{C_{\rm src}(15.6)}{C_{\rm src}(14)}
<\frac{3637.8947}{288}=12.6316.
\label{eq:source_cost_break_even}
\end{equation}
This multiplicative per-qubit break-even relation leaves source-specific shared hardware, nonlinear multiplexing overhead, and throughput constraints to a dedicated source model.

Equation~\eqref{eq:resource_cliff_ratio} compares the same nominal factory-consumption metric at different input qualities; it is not a ${\rm C_4}$/RHG cost quotient.

\subsubsection{Relation to actual expected retry consumption}
For comparison, let $C_{\rm att}$ include all costs of one complete candidate attempt, including its consumed child resources and any local preparation or verification resources. If successive attempts are independent and identically distributed, each has success probability $P_{\rm acc}>0$, and a failed attempt is discarded before restarting under the same rule, then
\begin{equation}
\mathbb E[C_{\rm delivered}]
=\frac{\mathbb E[C_{\rm att}]}{P_{\rm acc}}.
\label{eq:actual_retry_cost}
\end{equation}
This renewal identity follows by writing the mean total cost as the mean cost of the first attempt plus $(1-P_{\rm acc})$ times a fresh total cost. Cost and acceptance within one attempt need not be independent; the restarted process must obey the same distribution. Recycling, adaptive stopping, shared caches, and partial retries require their own ledger and state-dependent model. Substituting $q$ for $P_{\rm acc}$ in this single-candidate retry identity requires the corresponding acceptance assumption. The fixed-quota batch penalty $M/n$ above is a separate accounting relation.

We use $288$ as the declared nominal normalization rather than as a separately measured complete expected cost. Physical QND/SUM gate consumption requires separate local gate and ancilla entries rather than reuse of the GKP-only recursion. 
\subsubsection{Conventional Knill-type postselection}
\label{subsec:knill_hard}

For comparison, conventional hard-postselected constructions abort preparation when a verification flag is triggered~\cite{Knill2005,AGP2008Postselection}. This is a different acceptance policy from the numerical soft-verification mode used here; the pass probability below is not an omitted factor in that mode. The rigorous postselection threshold framework can be viewed as a recursive conditional-channel construction: data are allowed to interact with an ancillary resource only after the latter has passed the appropriate verification tests.

At a single verification level, let $A_L\in\{0,1\}$ be the level-$L$ retention indicator and let $E_L$ be the designated-output logical-failure event defined in Sec.~\ref{sec:interface_contract}. The hard-postselected logical-error probability is the conditional event probability
\begin{equation}
\Plog^{\rm hard}(L)=\Pr(E_L\mid A_L=1),
\label{eq:hard_conditional_channel}
\end{equation}
with acceptance probability
\begin{equation}
\Pacc^{\rm hard}(L)=\Pr(A_L=1).
\end{equation}
For a recursive resource factory in which a level-$(L{+}1)$ parent consumes $N_{\rm child}$ level-$L$ child resources, the hard-postselected acceptance probability has the schematic form
\begin{equation}
P_{\rm del}^{\rm hard}(L{+}1)
\simeq
P_{\rm ver}^{(L+1)}
\left[P_{\rm del}^{\rm hard}(L)\right]^{N_{\rm child}},
\label{eq:hard_recursive_delivery}
\end{equation}
where $P_{\rm ver}^{(L+1)}$ is the probability that the parent-level syndrome tests pass. When child resources are instead prepared with separate retries, the corresponding expected resource-cost recursion is
\begin{equation}
R_{L+1}^{\rm hard}
\simeq
\frac{N_{\rm child}R_L^{\rm hard}}{P_{\rm ver}^{(L+1)}}.
\label{eq:hard_recursive_resource}
\end{equation}
In the ${\rm C_4}$ Knill-type recursive verification gadget considered here, $N_{\rm child}=12$. Equations~\eqref{eq:hard_recursive_delivery} and \eqref{eq:hard_recursive_resource} make explicit why hard postselection can produce high conditional fidelity but poor throughput: the accepted channel is improved by conditioning, while the preparation overhead grows as the acceptance probability decreases. Related GKP schemes have likewise used analog information, tracking, and concatenation with higher-level codes to reduce logical errors or physical-resource overhead~\cite{Fukui2018TrackingQEC,FukuiMatsuuraMenicucci2023Concatenated}.

\subsection{Diagnostic scans and independent confirmation}\label{app:qualification_diagnostics}

After independent evaluation at $13$, $14$, $15$, and $16$ dB under the specified Jeffreys convention, a diagnostic scan selected one point for a new confirmation. Its designated-data failure counts per $10^5$ samples were
\begin{center}
\begin{tabular}{ccc}
\toprule
$s_{\rm GKP}$ (dB) & noise off & noise on\\
\midrule
15.2 & 2 & 4\\
15.4 & 1 & 2\\
15.6 & 0 & 0\\
15.8 & 0 & 0\\
\bottomrule
\end{tabular}
\end{center}
The scan was used only to identify $15.6$ dB as the first zero-observed-failure point on the evaluated $0.2$-dB grid. The $10^{-5}$ certificate is based on a separate fixed-sample confirmation with $10^6$ selected samples per condition, the decoder and operating point fixed before evaluation, and no parameter reselection. Its two condition-specific Clopper--Pearson results provide the focal qualification.

\begin{figure*}[t]
\centering
\includegraphics[width=0.94\textwidth]{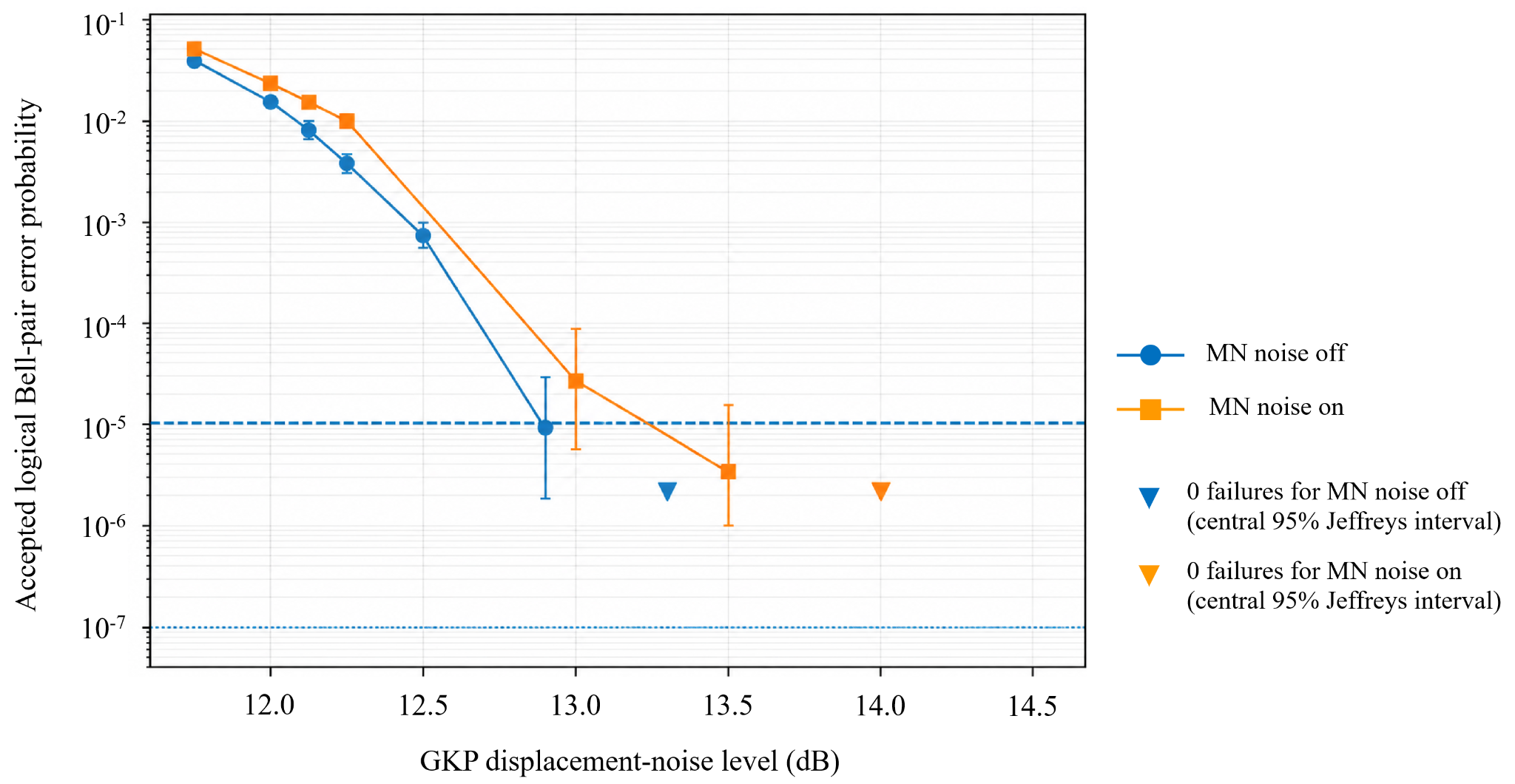}
\caption{Supplementary direct-tail diagnostics for the nominal level-3 retained fraction $q_3=0.75$. The displayed estimates and Jeffreys intervals refer to selected outputs; for the sealed zero-failure points, the plotted upper markers are the upper endpoints of the stored central $95\%$ Jeffreys credible intervals. Each sealed point contains $900{,}000$ selected outputs from $1{,}200{,}000$ raw trials. These adaptive-stopping diagnostics are not used as an additional independent fixed-sample target qualification. The $10^{-7}$ line is a reference scale only.}
\label{fig:direct_tail_validation}
\end{figure*}

The supplementary direct-tail calculation uses the nominal level-3 retained fraction $q_3=0.75$. The reported sealed diagnostic points have zero selected-output failures among $900{,}000$ selected outputs obtained from $1{,}200{,}000$ raw trials per condition. The stored central $95\%$ Jeffreys credible intervals have upper endpoints $2.794\times10^{-6}$ (MN noise off, $13.3$ dB) and $2.807\times10^{-6}$ (MN noise on, $14.0$ dB). These are Monte Carlo estimates of interval endpoints from $250{,}000$ posterior draws, not one-sided $95\%$ limits; their small difference at identical counts is sampling variation. Within each raw-data batch, candidates are ranked by observable risk and the prescribed lowest-risk $75\%$ is selected. This is not the application of a separately calibrated fixed cutoff. The adaptive-stopping diagnostic is therefore not an additional independent fixed-sample target qualification. Figure~\ref{fig:direct_tail_validation} is included only as a supplementary diagnostic; its $10^{-7}$ line is a reference scale, not a performance claim.

\subsection{C4-only plots and architecture boundaries}\label{app:rhg_parameters}
\begin{figure*}[t]
\centering
\includegraphics[width=0.83\textwidth]{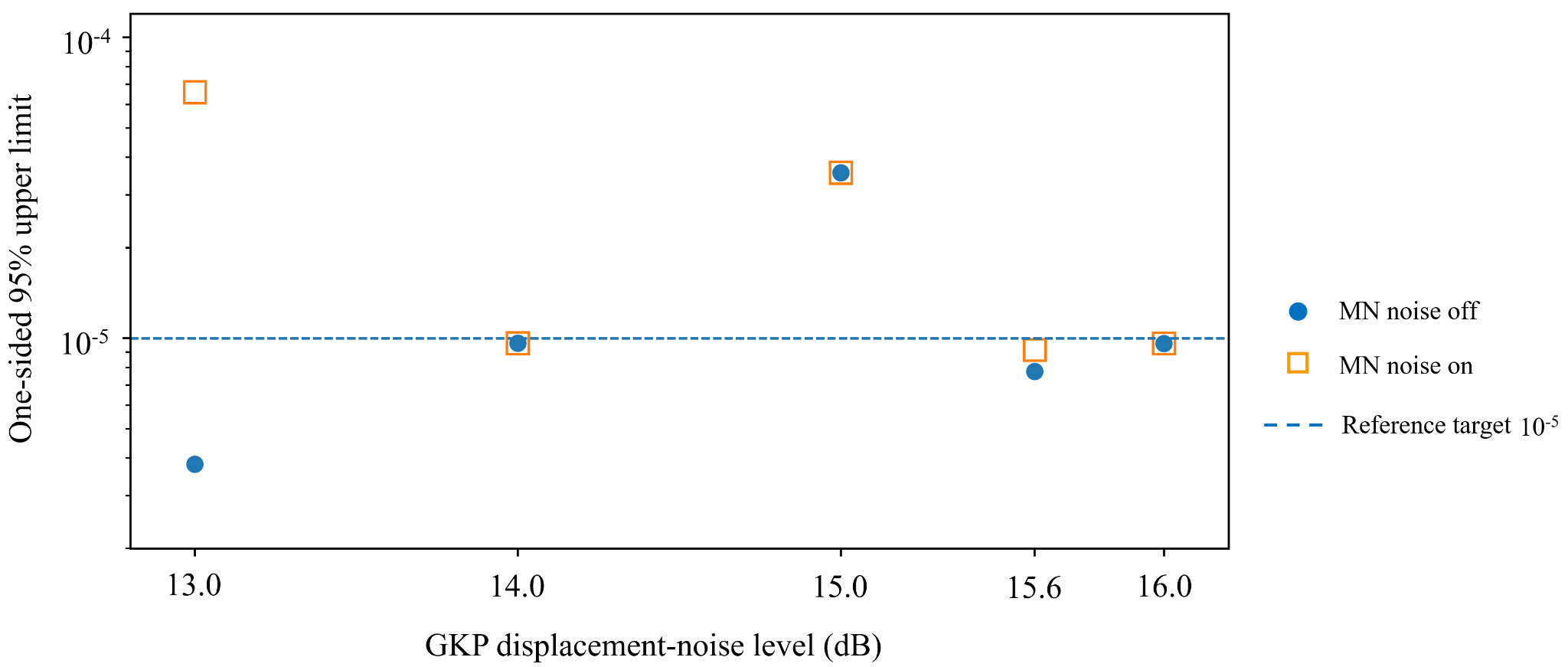}
\caption{${\rm C_4}$-only upper-bound arithmetic from the aggregate counts in Table~\ref{tab:qualified_operating_points}. The horizontal coordinate is the declared GKP displacement-noise level, not a direct measurement of a particular source. Filled circles and open squares distinguish MN noise off and on; overlapping markers denote equal upper bounds. The $15.6$-dB points use one-sided $95\%$ Clopper--Pearson limits, while the other points retain one-sided $95\%$ Jeffreys limits. Points are not joined because the evaluated level, retained fraction, and interval convention change. The limits are calculated from the reported $(k,n)$ under the statistical conventions and assumptions in Appendix~\ref{app:statistics}. Factory cost remains in Table~\ref{tab:qualified_operating_points}; the conditional SWAP-layer result is reported separately.}
\label{fig:system_level_synthesis}
\end{figure*}

The interface specification and within-factory comparison do not require a cross-architecture performance ranking. The nominal supplied-Bell-resource GKP count and an RHG spacetime volume have different units and are not directly comparable without matched source, operation, and failure conventions. Likewise, a one-layer logical SWAP construction is not a measurement of total algorithm runtime. Figure~\ref{fig:system_level_synthesis} shows the ${\rm C_4}$ aggregate-count upper bounds from Table~\ref{tab:qualified_operating_points}. It does not put a supplied Bell-resource cost and an RHG spacetime volume on one performance axis, or compare a one-layer SWAP construction with a different protected-operation duration.

A quantitative cross-architecture comparison would require the same logical operation, target failure definition, source convention, and finite-size model. The present comparison distinguishes these resource boundaries and does not establish a ${\rm C_4}$/RHG resource quotient, an optimal RHG distance, or a runtime advantage.

\clearpage
\section{QRL measurement patterns and their ${\rm C_4}$ logical action}
\label{app:qrl_logical_operations}
This Appendix gives representative QRL measurement patterns for the bounded-depth operations used at the ${\rm C_4}$/QRL interface and connects the physical Gaussian maps to the logical-action and protected-depth conventions of Sec.~\ref{subsec:interface_execution}. These ideal maps specify the physical implementation scope of the interface; the logical-error certificate for the accepted Bell resource is defined separately in the main text.

\subsection{Macronode and depth convention}
The physical macronode convention is the one shown in Fig.~\ref{MBQClayer}. The four measured modes in a macronode are denoted by $A$--$D$. In the diagrams below, quadrature labels attached to measured modes specify the homodyne bases, green circles mark the surviving computational rails, and the symbol $\circeq$ denotes equality of the induced Gaussian map up to deterministic classical postprocessing, displacement feedforward, and tracked Clifford-frame updates~\cite{walshe2020continuous,MenicucciQRL2020,Walshe2021Macronode}. These deterministic corrections belong to the correction record of Sec.~\ref{sec:interface_contract}, not additional protected quantum layers. An ideal Gaussian map in these diagrams denotes a Gaussian unitary, represented on quadratures by a symplectic transformation; its finite-squeezing noise channel is specified separately.

A protected online layer is one sequential QRL/macronode measurement layer acting on the injected encoded data. Physical operations on disjoint ${\rm C_4}$ indices belong to the same protected layer when they are executed in parallel. Thus, if a physical QRL pattern $g$ has fixed sequential length $d_g$, its transversal application to a level-$L$ block satisfies
\begin{equation}
D_{\rm prot}(g,L)=d_g
\end{equation}
under the advance-preparation and address-preserving routing conditions of Criterion~\ref{crit:distance_depth}. The present work does not assign separate numerical values to Hadamard, CNOT, or Bell-measurement depths; the quantitative benchmark used in the main text is the explicitly evaluated address-tracked logical-subsystem SWAP with $D_{\rm prot}^{\rm SWAP}=1$.

\subsection{Single-register identity and Fourier patterns}
The identity-teleportation pattern supplies the ideal single-step reference map. Its finite-squeezing calibration to the declared effective MN-noise condition is a separate requirement. A convenient homodyne choice is shown in Fig.~\ref{fig:qrl_identity_pattern}. Equal basis choices on the paired macronode modes cancel the unwanted vertical coupling after classical postprocessing, leaving the one-dimensional teleportation map~\cite{walshe2020continuous,MenicucciQRL2020}. Finite-squeezing noise associated with this step is treated separately in Appendix~\ref{sec:mbqc_insertion_noise}; the diagram below represents the ideal Gaussian map.

\par\medskip\noindent
\begin{minipage}{\columnwidth}
\makeatletter\def\@captype{figure}\makeatother
\centering
\resizebox{0.96\columnwidth}{!}{%
\begin{tikzpicture}[x=1cm,y=1cm]

  \newcommand{\configuration}[3]{%
    \begin{scope}[xshift=#1]
      \def\xOne{-1.655}
      \def\xTwo{0.245}
      \def\dx{0.36}

      #2

      \draw[horizontal rail]
        ({\xOne+\dx},0) -- ({\xTwo-\dx},0);
      \draw[horizontal rail]
        ({\xTwo+\dx},0) -- (1.295,0);

      \draw[vertical rail]
        (\xTwo,1.05) -- (\xTwo,\dx);
      \draw[vertical rail]
        (\xTwo,-\dx) -- (\xTwo,-1.05);

      \draw[cell arrow 2]
        ({\xTwo-\dx},0) to[bend left=44] (\xTwo,\dx);
      \draw[cell arrow 1]
        (\xTwo,\dx)
        .. controls ({\xTwo-.10},.20) and ({\xTwo-.10},-.20) ..
        (\xTwo,-\dx);
      \draw[cell arrow 2]
        (\xTwo,-\dx) to[bend right=44] ({\xTwo+\dx},0);
      \draw[cell arrow 1]
        ({\xTwo+\dx},0)
        .. controls ({\xTwo+.20},-.15) and ({\xTwo-.20},-.15) ..
        ({\xTwo-\dx},0);

      \node[pale green node] at ({\xOne+\dx},0) {};
      \node[pale yellow node] at ({\xTwo-\dx},0) {};
      \node[pale yellow node] at (\xTwo,\dx) {};
      \node[pale green node]  at ({\xTwo+\dx},0) {};
      \node[pale yellow node] at (\xTwo,-\dx) {};

      #3
    \end{scope}%
  }

  \configuration{-2.05cm}{}{%

    \node[selected label,anchor=south east]
      at ({\xTwo-.07},{\dx+.06}) {$\hat{q}_B$};
    \node[selected label,anchor=north east]
      at ({\xTwo-.04},{-\dx-.02}) {$\hat{p}_A$};
    \node[selected label,anchor=south east]
      at ({\xTwo-\dx-.05},.05) {$\hat{q}_C$};
    \node[selected label,anchor=south west]
      at ({\xTwo+\dx+.02},.03) {$\hat{p}_D$};
  }
  \node[font=\Large] at (0,0) {$\circeq$};
  \configuration{2.05cm}{%
    \path[connected highlight]
      ({\xOne+\dx-.24},-.24)
      rectangle
      ({\xTwo+\dx+.24},.24);
  }{%
    \node[selected label,anchor=south]
      at ({(\xOne+\xTwo)/2+\dx},.27) {{\red $\hat{I}$}};
  }

  \node[font=\sffamily\footnotesize,text=black!70] at (-2.05,1.35) {measurement pattern};
  \node[font=\sffamily\footnotesize,text=black!70] at (2.05,1.35) {effective map};
\end{tikzpicture}
}
\caption{Representative QRL identity-teleportation pattern. The left copy shows the homodyne bases and the right copy shows the induced ideal map. The equivalence is up to deterministic feedforward displacements and tracked frame updates.}
\label{fig:qrl_identity_pattern}
\end{minipage}
\par\medskip

A Fourier transformation is obtained by changing the measured quadratures as shown in Fig.~\ref{fig:qrl_fourier_pattern}~\cite{MenicucciQRL2020}. For square-lattice GKP qubits this physical Fourier map induces the usual logical Hadamard action up to the declared GKP/Clifford-frame convention. At the ${\rm C_4}$ level, however, the label-preserving Hadamard also requires the physical-index permutation specified below; this is why the logical action is stated in terms of the ${\rm C_4}$ $D/G$ subsystem map rather than by assigning a separate depth symbol to the Fourier pattern alone.

\par\medskip\noindent
\begin{minipage}{\columnwidth}
\makeatletter\def\@captype{figure}\makeatother
\centering
\resizebox{0.96\columnwidth}{!}{%
\begin{tikzpicture}[x=1cm,y=1cm]

  \newcommand{\configuration}[3]{%
    \begin{scope}[xshift=#1]
      \def\xOne{-1.655}
      \def\xTwo{0.245}
      \def\dx{0.36}

      #2

      \draw[horizontal rail]
        ({\xOne+\dx},0) -- ({\xTwo-\dx},0);
      \draw[horizontal rail]
        ({\xTwo+\dx},0) -- (1.295,0);

      \draw[vertical rail]
        (\xTwo,1.05) -- (\xTwo,\dx);
      \draw[vertical rail]
        (\xTwo,-\dx) -- (\xTwo,-1.05);

      \draw[cell arrow 2]
        ({\xTwo-\dx},0) to[bend left=44] (\xTwo,\dx);
      \draw[cell arrow 1]
        (\xTwo,\dx)
        .. controls ({\xTwo-.10},.20) and ({\xTwo-.10},-.20) ..
        (\xTwo,-\dx);
      \draw[cell arrow 2]
        (\xTwo,-\dx) to[bend right=44] ({\xTwo+\dx},0);
      \draw[cell arrow 1]
        ({\xTwo+\dx},0)
        .. controls ({\xTwo+.20},-.15) and ({\xTwo-.20},-.15) ..
        ({\xTwo-\dx},0);

      \node[pale green node] at ({\xOne+\dx},0) {};
      \node[pale yellow node] at ({\xTwo-\dx},0) {};
      \node[pale yellow node] at (\xTwo,\dx) {};
      \node[pale green node]  at ({\xTwo+\dx},0) {};
      \node[pale yellow node] at (\xTwo,-\dx) {};

      #3
    \end{scope}%
  }

  \configuration{-2.35cm}{}{%

    \node[selected label,anchor=south east]
      at ({\xTwo-.07},{\dx+.06}) {$(\hat{q} + \hat{p})_B$};
    \node[selected label,anchor=north east]
      at ({\xTwo-.04},{-\dx-.02}) {$(\hat{p}-\hat{q})_A$};
    \node[selected label,anchor=south east]
      at ({\xTwo-\dx-.05},.05) {$(\hat{q} + \hat{p})_C$};
    \node[selected label,anchor=south west]
      at ({\xTwo+\dx+.02},.03) {$(\hat{p}-\hat{q})_D$};
  }
  \node[font=\Large] at (0,0) {$\circeq$};
  \configuration{2.35cm}{%
    \path[connected highlight]
      ({\xOne+\dx-.24},-.24)
      rectangle
      ({\xTwo+\dx+.24},.24);
  }{%
    \node[selected label,anchor=south]
      at ({(\xOne+\xTwo)/2+\dx},.27) {{\red $\hat{F}$}};
  }

  \node[font=\sffamily\footnotesize,text=black!70] at (-2.35,1.35) {measurement pattern};
  \node[font=\sffamily\footnotesize,text=black!70] at (2.35,1.35) {effective map};
\end{tikzpicture}
}
\caption{Representative QRL Fourier pattern. The highlighted effective map is $\hat F$ up to deterministic displacement and frame corrections. Parallel application to the ${\rm C_4}$ leaves supplies the physical $H^{\otimes4}$ component of the label-preserving logical Hadamard.}
\label{fig:qrl_fourier_pattern}
\end{minipage}
\par\medskip

\subsection{Tracked register permutation and the SWAP benchmark}
QRL measurement choices can also implement deterministic permutations of the computational registers. Figure~\ref{fig:qrl_swap_pattern} shows a two-register SWAP pattern. The operation moves the computational rails while exchanging the selected registers; for some larger cyclic permutations a known Fourier byproduct can appear and must be retained in the frame record. Because the permutation is deterministic, it is propagated through the declared address map rather than treated as an unknown error.

For the ${\rm C_4}$ logical-subsystem SWAP used as the depth benchmark, the required physical permutation is applied in parallel across the relevant leaves and the output-address record is relabeled to follow the routed data subsystem. The specific benchmark used in the main text is one protected macronode layer,
\begin{equation}
D_{\rm prot}^{\rm SWAP}=1,
\end{equation}
while controller latency and frame bookkeeping are accounted for separately.

\subsection{Entangling SUM--SWAP pattern and ${\rm C_4}$ logical action}
The two-register entangling pattern used here is a SUM operation accompanied by a deterministic register exchange. The corresponding QRL measurement pattern is shown in Fig.~\ref{fig:qrl_sumswap_pattern}. In the vertical crossing channel, the sequence of local Fourier transformations and the QRL entangling interaction produces the physical SUM map; the same pattern also exchanges the routed computational registers. Thus the physical operation furnished directly by this pattern is a SUM--SWAP rather than a bare CNOT with an implicit routing convention.

On square-lattice GKP code states, the SUM map induces the logical GKP CX/CNOT action up to the declared local Fourier and Pauli/Clifford-frame conventions. Applied in parallel to corresponding physical indices of two ${\rm C_4}$ blocks, the SUM component therefore supplies the transversal logical CNOT action, while the accompanying deterministic register exchange is propagated through the address record. In the ${\rm C_4}$ logical-Pauli convention of Appendix~\ref{app:c4_two_logical_subsystems}, the abstract transversal CNOT map is
\begin{equation}
\operatorname{CNOT}^{\otimes4}_{A\rightarrow B}
=
\overline{\operatorname{CNOT}}_{D_A\rightarrow D_B}
\otimes
\overline{\operatorname{CNOT}}_{G_A\rightarrow G_B},
\end{equation}
and the physical QRL realization adds only the declared deterministic routing/permutation shown in Fig.~\ref{fig:qrl_sumswap_pattern}. The quantitative depth comparison in this work is anchored by the SWAP benchmark and therefore does not require an additional operation-specific CNOT depth parameter in the numerical claim.

\par\medskip\noindent
\begin{minipage}{\columnwidth}
\makeatletter\def\@captype{figure}\makeatother
\centering
\resizebox{0.96\columnwidth}{!}{%
\begin{tikzpicture}[x=1cm,y=1cm]

  \newcommand{\configuration}[4]{%
    \begin{scope}[xshift=#1]
      \def\xA{-1.655}
      \def\xB{0.245}
      \def\dx{0.36}

      #3

      \foreach \y in {1.90,0,-1.90} {
        \draw[horizontal rail]
          ({\xA+\dx},\y) -- ({\xB-\dx},\y);
        \draw[horizontal rail]
          ({\xB+\dx},\y) -- (1.295,\y);
      }

      \draw[vertical rail] (\xB,2.54) -- (\xB,2.26);
      \draw[vertical rail] (\xB,1.54) -- (\xB,.36);
      \draw[vertical rail] (\xB,-.36) -- (\xB,-1.54);
      \draw[vertical rail] (\xB,-2.26) -- (\xB,-2.54);

      \foreach \y/\r/\aLT/\aTB/\aBR/\aRL in {
   1.90/0/1/2/1/2,%
   0/1/2/1/2/1,%
  -1.90/2/1/2/1/2%
} {
  \coordinate (l#2\r) at ({\xB-\dx},\y);
  \coordinate (t#2\r) at (\xB,{\y+\dx});
  \coordinate (r#2\r) at ({\xB+\dx},\y);
  \coordinate (b#2\r) at (\xB,{\y-\dx});

  \draw[cell arrow \aLT]
    (l#2\r) to[bend left=44] (t#2\r);

  \draw[cell arrow \aTB]
    (t#2\r)
    .. controls ({\xB-.10},{\y+.20})
       and ({\xB-.10},{\y-.20}) ..
    (b#2\r);

  \draw[cell arrow \aBR]
    (b#2\r) to[bend right=44] (r#2\r);

  \draw[cell arrow \aRL]
    (r#2\r)
    .. controls ({\xB+.20},{\y-.15})
       and ({\xB-.20},{\y-.15}) ..
    (l#2\r);
}

      \foreach \y/\r in {1.90/0,0/1,-1.90/2} {
        \node[pale green node] at ({\xA+\dx},\y) {};
        \node[pale yellow node] at (l#2\r) {};
        \node[pale yellow node] at (t#2\r) {};
        \node[pale green node]  at (r#2\r) {};
        \node[pale yellow node] at (b#2\r) {};
      }

      #4
    \end{scope}%
  }

  \newcommand{\placecirclelabels}[5]{%
    \node[selected label,anchor=south east]
      at ({\xB-.07},{#1+\dx+.07}) {#3};
    \node[selected label,anchor=north east]
      at ({\xB+.02},{#1-\dx+.02}) {#2};
    \node[selected label,anchor=south east]
      at ({\xB-\dx-.05},{#1+.05}) {#4};
    \node[selected label,anchor=south west]
      at ({\xB+\dx+.05},{#1+.05}) {#5};
  }

  \newcommand{\circlelabels}{%
    \placecirclelabels{1.90}
      {\(\hat{q}_{A_1}\)}
      {\(\hat{q}_{B_1}\)}
      {\(\hat{p}_{C_1}\)}
      {\(\hat{p}_{D_1}\)}
    \placecirclelabels{0}
      {\(\hat{p}_{A_2}\)}
      {\(\hat{q}_{B_2}\)}
      {\(\hat{q}_{C_2}\)}
      {\(\hat{p}_{D_2}\)}
    \placecirclelabels{-1.90}
      {\(\hat{q}_{A_3}\)}
      {\(\hat{q}_{B_3}\)}
      {\(\hat{p}_{C_3}\)}
      {\(\hat{p}_{D_3}\)}
  }

  \newcommand{\regionlabels}{%
    \node[selected label,anchor=east] at ({\xB-.32},.95) {};
    \node[selected label,anchor=east] at ({\xB-.32},-.95) {\red SWAP};
  }

  \configuration{-2.05cm}{L}{}{\circlelabels}
  \node[font=\Large] at (0,0) {$\circeq $};
  \configuration{2.05cm}{R}{%

    \draw[zigzag region]
      ({\xB+\dx},1.90)
      -- (\xB,{1.90-\dx})
      -- (\xB,0)
      -- (\xB,{-1.90+\dx})
      -- ({\xB-\dx},-1.90)
      -- ({\xA+\dx},-1.90);
    \draw[zigzag region]
      ({\xA+\dx},0)
      -- ({\xB+\dx},0);
  }{\regionlabels}

  \node[font=\sffamily\footnotesize,text=black!70] at (-2.05,2.80) {measurement pattern};
  \node[font=\sffamily\footnotesize,text=black!70] at (2.05,2.80) {tracked register map};
\end{tikzpicture}
}
\caption{QRL register-permutation pattern implementing a deterministic SWAP of the selected computational registers, up to tracked frame corrections. The right-hand shaded path denotes the effective routed register map. The permutation is retained explicitly in the address record and is therefore not interpreted as a stochastic logical fault.}
\label{fig:qrl_swap_pattern}
\end{minipage}
\par\medskip

\par\medskip\noindent
\begin{minipage}{\columnwidth}
\makeatletter\def\@captype{figure}\makeatother
\centering
\resizebox{0.96\columnwidth}{!}{%
\begin{tikzpicture}[x=1cm,y=1cm]

  \newcommand{\configuration}[4]{%
    \begin{scope}[xshift=#1]
      \def\xA{-1.655}
      \def\xB{0.245}
      \def\dx{0.36}

      #3

      \foreach \y in {1.90,0,-1.90} {
        \draw[horizontal rail]
          ({\xA+\dx},\y) -- ({\xB-\dx},\y);
        \draw[horizontal rail]
          ({\xB+\dx},\y) -- (1.295,\y);
      }

      \draw[vertical rail] (\xB,2.54) -- (\xB,2.26);
      \draw[vertical rail] (\xB,1.54) -- (\xB,.36);
      \draw[vertical rail] (\xB,-.36) -- (\xB,-1.54);
      \draw[vertical rail] (\xB,-2.26) -- (\xB,-2.54);

      \foreach \y/\r/\aLT/\aTB/\aBR/\aRL in {
   1.90/0/1/2/1/2,%
   0/1/2/1/2/1,%
  -1.90/2/1/2/1/2%
} {
  \coordinate (l#2\r) at ({\xB-\dx},\y);
  \coordinate (t#2\r) at (\xB,{\y+\dx});
  \coordinate (r#2\r) at ({\xB+\dx},\y);
  \coordinate (b#2\r) at (\xB,{\y-\dx});

  \draw[cell arrow \aLT]
    (l#2\r) to[bend left=44] (t#2\r);

  \draw[cell arrow \aTB]
    (t#2\r)
    .. controls ({\xB-.20},{\y+.22})
       and ({\xB-.10},{\y-.20}) ..
    (b#2\r);

  \draw[cell arrow \aBR]
    (b#2\r) to[bend right=44] (r#2\r);

  \draw[cell arrow \aRL]
    (r#2\r)
    .. controls ({\xB+.20},{\y-.15})
       and ({\xB-.20},{\y-.15}) ..
    (l#2\r);
}

      \foreach \y/\r in {1.90/0,0/1,-1.90/2} {
        \node[pale green node] at ({\xA+\dx},\y) {};
        \node[pale yellow node] at (l#2\r) {};
        \node[pale yellow node] at (t#2\r) {};
        \node[pale green node]  at (r#2\r) {};
        \node[pale yellow node] at (b#2\r) {};
      }

      #4
    \end{scope}%
  }

  \newcommand{\placecirclelabels}[5]{%
    \node[selected label,anchor=south east]
      at ({\xB-.07},{#1+\dx+.07}) {#3};
    \node[selected label,anchor=north east]
      at ({\xB+.02},{#1-\dx+.05}) {#2};
    \node[selected label,anchor=south east]
      at ({\xB-\dx-.05},{#1+.05}) {#4};
    \node[selected label,anchor=south west]
      at ({\xB+\dx+.01},{#1+.05}) {#5};
  }

  \newcommand{\circlelabels}{%
    \placecirclelabels{1.90}
      {\(\hat{q}_{A_1}\)}
      {\(\hat{q}_{B_1}\)}
      {\(\hat{p}_{C_1}\)}
      {\(\hat{p}_{D_1}\)}
    \placecirclelabels{0}
      {\((\hat{p}-\frac{\hat{q}}{2})_{A_2}\)}
      {\(\hat{q}_{B_2}\)}
      {\(\hat{q}_{C_2}\)}
      {\((\hat{p}+\frac{\hat{q}}{2})_{D_2}\)}
    \placecirclelabels{-1.90}
      {\(\hat{q}_{A_3}\)}
      {\(\hat{q}_{B_3}\)}
      {\(\hat{p}_{C_3}\)}
      {\(\hat{p}_{D_3}\)}
  }

  \newcommand{\regionlabels}{%
    \node[selected label,anchor=east] at ({\xB-.32},.95) {};
    \begin{scope}[
  shift={({\xB-.8},-1.1)},
  draw=red!70!black,
  line width=.7pt
]
  
  \draw (0,.5) -- (0,-.12);

  \fill[red!70!black] (0,.5) circle[radius=1.5pt];

  \draw (0,-.12) circle[radius=4pt];
  \draw (-.12,-.12) -- (.12,-.12);
  \draw (0,-.24) -- (0,-.04);
\end{scope}
}

  \configuration{-2.45cm}{L}{}{\circlelabels}
  \node[font=\Large] at (0,0) {$\circeq $};
  \configuration{2.45cm}{R}{%

    \draw[zigzag region]
      ({\xB+\dx},1.90)
      -- (\xB,{1.90-\dx})
      -- (\xB,0)
      -- (\xB,{-1.90+\dx})
      -- ({\xB-\dx},-1.90)
      -- ({\xA+\dx},-1.90);
    \draw[zigzag region]
      ({\xA+\dx},0)
      -- ({\xB+\dx},0);
  }{\regionlabels}

  \node[font=\sffamily\footnotesize,text=black!70] at (-2.45,2.80) {measurement pattern};
  \node[font=\sffamily\footnotesize,text=black!70] at (2.45,2.80) {SUM + tracked register map};
\end{tikzpicture}
}
\caption{QRL SUM--SWAP measurement pattern. The chosen homodyne bases induce the physical SUM entangling map together with a deterministic exchange of the computational registers, up to displacement and frame corrections. When applied transversally to address-matched GKP/${\rm C_4}$ leaves, the SUM component supplies the logical CNOT action and the known register exchange is carried by the address record.}
\label{fig:qrl_sumswap_pattern}
\end{minipage}
\par\medskip

\subsection{${\rm C_4}$ Hadamard, Bell measurement, and non-Clifford injection}
For the data and gauge degrees of freedom of ${\rm C_4}$, the label-preserving Hadamard convention used throughout the manuscript is
\begin{equation}
\overline H_D\overline H_G
=
\operatorname{SWAP}_{23}H^{\otimes4},
\end{equation}
consistent with Eq.~\eqref{eq:label_preserving_H}. The parallel Fourier pattern of Fig.~\ref{fig:qrl_fourier_pattern} supplies the $H^{\otimes4}$ component, while the required deterministic physical-index permutation is implemented and tracked using the register-routing operation above. The purpose of this construction is to establish a bounded-length physical realization consistent with the interface criterion; no separate Hadamard depth parameter enters the numerical claim.

Logical Bell measurement is implemented by parallel physical GKP Bell measurements followed by classical ${\rm C_4}$ decoding and deterministic frame composition. Only the physical measurement sequence contributes to protected quantum depth; the decoding and frame-processing latency belongs to the classical-control term of the latency decomposition. The numerical logical-error certificate is defined for the accepted encoded Bell resource and the declared post-injection channel; the ideal maps in this Appendix specify how the corresponding QRL operations enter the interface.

Fault-tolerant universality additionally requires a verified non-Clifford resource. Such a resource can be prepared offline and consumed through logical teleportation, with the resulting Pauli/Clifford byproduct propagated through the same frame and address interface~\cite{BravyiKitaev2005Magic,BravyiHaah2012Magic}. The architecture is compatible with this route, but the numerical resource counts $288$ and $3637.9$ reported in the main text do not include a non-Clifford-resource factory, and no full algorithm-level compilation or waiting-time claim is made here.
\subsection{Advance preparation and the execution criterion}\label{app:execution_criterion}
$D_{\rm prot}$ counts protected QRL/macronode layers, not preparation, classical processing, or resource waiting. These circuit-level counts are distinct from the algorithm-level guarantees of transversal algorithmic fault tolerance~\cite{Zhou2025TransversalAFT}.

\begin{criterionbox}[Preservation of bounded online depth under advance preparation and parallel routing]
\label{crit:distance_depth}
Consider a level-$L$ code family and a logical operation $G$ for which there exists a level-independent integer $D_G$ such that, for every $L$, $G$ is implemented by $D_G$ sequential layers of disjoint physical operations. Suppose that (i) every encoded ancilla required by $G$ is prepared before the online operation, (ii) all physical operations within each declared layer are executable concurrently with the available parallel capacity, (iii) routing preserves that parallel execution without serializing a declared protected layer, and (iv) classical Pauli-frame updates are excluded from protected online depth. Then
\begin{equation}
D_{\rm prot}(G,L)=D_G,
\label{eq:bounded_online_depth}
\end{equation}
independent of $L$, even when the code distance $d(L)$ and the number of parallel physical operations $N_{\rm pair}(L)$ increase with $L$. For concatenated ${\rm C_4}$, the code distance is $d_{\rm C4}(L)=2^L$ and $N_{\rm pair}(L)=4^L=d_{\rm C4}(L)^2$.
\end{criterionbox}

Advance preparation does not imply zero or level-independent preparation time. Replacing logical locations by disjoint child locations increases parallel width without adding protected layers only when each layer can run concurrently. Hardware serialization violates this premise and must be accounted for through $B_{\rm addr}$ and the timing model below. This is an implementation criterion; finite-level error contraction is evaluated separately.

\subsection{Protected online depth, parallel width, and execution time}\label{subsec:online_timing}
For a level-$L$ ${\rm C_4}$ block, the number of physical operations required by one transversal layer is
\begin{equation}
N_{\rm pair}(L)=4^L=d_{\rm C4}(L)^2.
\label{eq:pairing_width}
\end{equation}
If only $P_{\rm pair}$ operations are available in one optical clock slot, the transversal layer spans
\begin{equation}
B_{\rm addr}(L;P_{\rm pair})=
\left\lceil\frac{4^L}{P_{\rm pair}}\right\rceil
\label{eq:address_span}
\end{equation}
time bins or synchronized batches. Criterion~\ref{crit:distance_depth} concerns protected online depth, not total elapsed time. A schematic timing relation is
\begin{equation}
\tau_{\rm op}\gtrsim B_{\rm addr}\tau_{\rm bin}+D_G\tau_{\rm MN}+\tau_{\rm ff},
\label{eq:operation_latency_decomposition}
\end{equation}
where $\tau_{\rm bin}$ is the time-bin spacing, $\tau_{\rm MN}$ is the duration of one QRL/macronode layer, and $\tau_{\rm ff}$ is the feedforward latency. Full spatial parallelism gives $B_{\rm addr}=1$; temporal multiplexing reduces the number of spatial modes at the cost of a longer operation time. First-output latency also includes resource preparation and availability, for which no distance-independent bound is established.

Let $\Gamma_{\rm GKP}(s)$ denote the rate at which physical GKP qubits at displacement-noise level $s$ enter the factory, and let $R_{\rm GKP,L}(s)$ be the expected supplied-qubit consumption per accepted encoded Bell resource. When $R_{\rm GKP,L}$ is the actual mean delivered-resource cost and GKP supply is the bottleneck, the ideal source-limited accepted-resource rate is
\begin{equation}
\Gamma_{\rm fac,L}(s)\simeq
\frac{\Gamma_{\rm GKP}(s)}{R_{\rm GKP,L}(s)}.
\label{eq:source_limited_factory_rate}
\end{equation}
Inserting the nominal schedule cost into this relation is a conditional planning estimate, not a measured throughput. It additionally requires that the nominal baseline and retention penalties equal the corresponding actual mean costs.
If a logical operation $G$ consumes $n_{\rm anc}(G)$ accepted resources at rate $\Gamma_G$, stall-free operation requires
\begin{equation}
\Gamma_{\rm fac,L}(s)\ge n_{\rm anc}(G)\Gamma_G.
\label{eq:stall_free_condition}
\end{equation}
This average-rate requirement alone does not exclude stalls with stochastic supply and finite buffers.
Within this source-limited model, assigning equal raw GKP supply rates at $14$ and $15.6$ dB would make the accepted-resource rate at the $L=2$ point a factor $3637.8947/288=12.6316$ higher than at the $L=3$ point. This conditional rate estimate assumes that entangling gates, switching, storage, and decoding are not bottlenecks; it is not a measured speedup. Calibrated storage loss and finite-capacity throughput are outside the present numerical evaluation.

The two operating points highlighted in the resource comparison are
\begin{align}
L=2:&\quad d_{\rm C4}(2)=4,\quad N_{\rm pair}=16,\quad D_{\rm prot}^{\rm SWAP}=1,\nonumber\\[-2pt]
&\quad R_{\rm GKP}=288\quad (15.6\ \mathrm{dB}),\nonumber\\[2pt]
L=3:&\quad d_{\rm C4}(3)=8,\quad N_{\rm pair}=64,\quad D_{\rm prot}^{\rm SWAP}=1,\nonumber\\[-2pt]
&\quad R_{\rm GKP}=3637.9\quad (14\ \mathrm{dB}).
\label{eq:distance_depth_operating_points}
\end{align}
These points illustrate two uses of the same interface at different input qualities, block sizes, and factory costs. Their one-layer SWAP construction is a separate conditional circuit property, not a fixed-quality asymptotic fit or a total-runtime comparison.

For implementation planning, the classical feedforward loop may be summarized by
\begin{equation}
B_{\rm delay}=\left\lceil \frac{\tau_{\rm ff}}{\tau_{\rm bin}}\right\rceil,
\end{equation}
where $\tau_{\rm bin}$ is the optical time-bin spacing and $\tau_{\rm ff}$ is the feedforward latency. An algorithm-level estimate would additionally require a compiled logical circuit, counts and scheduling of Clifford and non-Clifford operations, an allocated total failure budget, the refresh/teleportation cadence, magic-state-factory throughput, routing, storage, and concurrent factory/processor timing. Because these inputs are not fixed here, we do not report an algorithm-level runtime or spacetime volume.

\bibliography{references}

\end{document}